\documentclass[english]{IEEEtran}
\usepackage[T1]{fontenc}
\usepackage[latin9]{inputenc}
\usepackage{color}
\usepackage{float}
\usepackage{amsmath}
\usepackage{amssymb}
\usepackage{graphicx}

\makeatletter

\floatstyle{ruled}
\newfloat{algorithm}{tbp}{loa}
\providecommand{\algorithmname}{Algorithm}
\floatname{algorithm}{\protect\algorithmname}

\usepackage{flushend}

\@ifundefined{showcaptionsetup}{}{%
 \PassOptionsToPackage{caption=false}{subfig}}
\usepackage{subfig}
\makeatother

\usepackage{babel}
\begin{document}
\title{On the Complexity Reduction of Uplink Sparse Code Multiple Access
for Spatial Modulation}
\author{Ibrahim Al-Nahhal, \textit{Student Member, IEEE},\textit{ }Octavia
A. Dobre, \textit{Fellow, IEEE}, and\\
Salama Ikki, \textit{Senior} \textit{Member, IEEE }\thanks{This work is supported by the Natural Sciences and Engineering Research
Council of Canada (NSERC), through its Discovery program.}\thanks{I. Al-Nahhal are with the Faculty of Engineering and Applied Science,
Memorial University, St. John\textquoteright s, NL, Canada, and is
also on the leave of the Faculty of Engineering, Al-Azhar University,
Cairo, Egypt (e-mail: ioalnahhal@mun.ca).}\thanks{O. A. Dobre is with the Faculty of Engineering and Applied Science,
Memorial University, St. John\textquoteright s, NL, Canada (e-mail:
odobre@mun.ca).}\thanks{S. Ikki is with the Department of Electrical Engineering, Lakehead
University, Thunder Bay, ON, Canada (e-mail: sikki@lakeheadu.ca).}}
\maketitle
\begin{abstract}
Multi-user spatial modulation (SM) assisted by sparse code multiple
access (SCMA) has been recently proposed to provide uplink high spectral
efficiency transmission. The message passing algorithm (MPA) is employed
to detect the transmitted signals, which suffers from high complexity.
This paper proposes three low-complexity algorithms for the first
time to the SM-SCMA. The first algorithm is referred to as successive
user detection (SUD), while the second algorithm is the modified version
of SUD, namely modified SUD (MSUD). Then, for the first time, the
tree-search of the SM-SCMA is constructed. Based on that tree-search,
another variant of the sphere decoder (SD) is proposed for the SM-SCMA,
referred to as fixed-complexity SD (FCSD). SUD provides a benchmark
for decoding complexity at the expense of bit-error-rate (BER) performance.
Further, MSUD slightly increases the complexity of SUD with a significant
improvement in BER performance. Finally, FCSD provides a near-optimum
BER with a considerable reduction of the complexity compared to the
MPA decoder and also supports parallel hardware implementation. The
proposed algorithms provide flexible design choices for practical
implementation based on system design demands. The complexity analysis
and Monte-Carlo simulations of the BER are provided for the proposed
algorithms.
\end{abstract}


\section{Introduction}

\def\figurename{Fig.}
\def\tablename{TABLE}

\IEEEPARstart{N}{on-orthogonal} multiple access (NOMA) has been recognized
as a promising technique for future wireless networks, and has received
considerable attention in recent years {[}\ref{W.-Shin,-M.}{]}-{[}\ref{A. Yadav}{]}.
NOMA is composed of two types: power-domain and code-domain. The power
and code orthogonality constraints are relaxed for multiple-user access
to improve the spectral efficiency and increase the number of served
users for power-domain and code-domain NOMA, respectively {[}\ref{M.-Mohammadkarimi,-M.}{]}-{[}\ref{Z.-Ding-et}{]}.
In this paper, sparse code multiple access (SCMA) code-domain NOMA
is considered, which was firstly proposed in {[}\ref{H.-Nikopour-and}{]}.
In the SCMA scheme, a unique multidimensional codebook is assigned
to each user to share the medium with the other users. The SCMA codebooks
are sparse (i.e., contain zeros) and carefully designed to provide
a good performance {[}\ref{M.-Taherzadeh-et}{]}-{[}\ref{M.-Vameghestahbanati,-I.}{]}.
The sparsity property of the SCMA codebooks makes it feasible to employ
the iterative message passing algorithm (MPA) to provide near maximum-likelihood
(ML) bit-error-rate (BER) performance at low-complexity detection
{[}\ref{H.-Mu,-Z.}{]}. The complexity of the MPA is still high for
practical implementations; several algorithms have been proposed to
tackle this problem {[}\ref{L.-Yang,-Y.}{]}-{[}\ref{L.-Li,-J. Modified SD}{]}.
In {[}\ref{L.-Yang,-Y.}{]}, the authors proposed a reduced-complexity
version of the MPA decoder, whereas the authors in {[}\ref{J.-Dai,-K.}{]}-{[}\ref{L.-Li,-J. Modified SD}{]}
adopted the concept of the sphere decoder (SD) to reduce the complexity
of the SCMA signal detection.

On the other hand, spatial modulation (SM) is a promising technology
for single-user communications, which overcomes the inter-channel-interference
problem present in multiple-input multiple-output (MIMO) schemes and
uses a single radio frequency (RF) chain {[}\ref{R.-Y.-Mesleh}{]}-{[}\ref{E.-Basar,-=002018=002018Index}{]}.
Since most of the existing systems already contain multiple antennas
at both transmitter and receiver, the SM system becomes a promising
candidate for those applications that cannot afford the aforementioned
MIMO drawbacks {[}\ref{M.-Renzo,-H.}{]}-{[}\ref{M.-Wen-et}{]}. The
SM system employs the index of the active antenna to deliver additional
information supplementary to the modulated quadrature amplitude modulation
(QAM)/phase-shift-keying (PSK) symbol that can be transmitted from
that active antenna {[}\ref{R.-Y.-Mesleh}{]}. At the receiver side,
the ML jointly detects the active transmit antenna as well as the
transmitted QAM/PSK symbol by implementing an exhaustive search that
leads to high decoding complexity. The algorithms in {[}\ref{A.-Younis,-R. GLOBECOM}{]}-{[}\ref{I.-Al-Nahhal,-E. JSAC}{]}
have been proposed based on the SD and tree-search concepts to significantly
reduce the decoding complexity of the SM system while retaining the
same BER performance of the ML decoder.

Since the single-user SM has recently been an attractive area of research,
it is vital to investigate the multi-user SM scenario using one of
the promising multiple access techniques, such as SCMA. Recently,
the multi-user SM has been assisted by SCMA (SM-SCMA) to provide a
high spectral efficiency transmission for uplink scenario {[}\ref{C.-Zhong,-X.}{]}-{[}\ref{Z.-Pan,-J.}{]}.
The SM-SCMA system requires a high number of transmit antennas to
provide high spectral efficiency for all users. To effectively tackle
this problem, the rotational generalized SM (RGSM)-SCMA has been proposed
in {[}\ref{I.-Al-Nahhal,-O. TVT}{]}. In the RGSM-SCMA system, the
same spectral efficiency of the SM-SCMA can be achieved using a significantly
reduced number of transmit antennas at the expense of almost negligible
changes to BER performance and decoding complexity, when compared
with the SM-SCMA system. For the SM-SCMA and RGSM-SCMA systems, the
iterative MPA decoder has been proposed to detect the transmitted
signal {[}\ref{Z.-Pan,-J.}{]}, {[}\ref{I.-Al-Nahhal,-O. TVT}{]}.
The MPA decoder iteratively updates the users message probabilities
until achieving the maximum number of iterations; this leads to an
increase in the decoding complexity of both systems. To the best of
the authors' knowledge, the MPA is the only existing decoder for the
SM-SCMA system.

In this paper, three low-complexity decoding algorithms for the uplink
SM-SCMA system are proposed. The first algorithm is termed successive
user detection (SUD). It detects the users messages that share the
first orthogonal resource element (ORE), then by using those detected
users messages, it successively detects the users messages that share
the next OREs. The SUD algorithm detects the user message using only
one of the available OREs that carry the signal of that user. The
proposed SUD algorithm is considered to be the lower bound of the
decoding complexity for the SM-SCMA and RGSM-SCMA systems at the expense
of the BER performance. By exploiting all available OREs for each
user with some iterative procedure, the modified SUD (MSUD) provides
a considerable improvement in the BER performance at the expense of
a small increase in the decoding complexity.

The SD and tree-search concepts are carefully designed for the uplink
SM-SCMA, referred to as a fixed-complexity SD (FCSD) algorithm. The
FCSD algorithm provides almost the same BER performance as that of
MPA with a significant reduction in the decoding complexity. The proposed
FCSD has a fixed decoding complexity for all values for signal-to-noise
ratio (SNR), as well as for its feasibility of parallel hardware implementation,
which is proper for practical applications {[}\ref{L.-G.-Barbero}{]},
{[}\ref{I.-Al-Nahhal,-M. Kbest ill condition}{]}. Besides, the FCSD
algorithm provides a favorable trade-off between the decoding complexity
and BER performance, which fits a wide range of practical applications.
In summary, each of the three proposed algorithms enjoys different
advantages that can fit a wide range of system specifications. The
complexity analysis in terms of the number of real additions and multiplications
is derived. The Monte-Carlo simulations for the BER performance of
the proposed algorithms are provided to support the paper findings.

The summary of the paper contributions is as follows:
\begin{enumerate}
\item Propose a benchmark low-complexity decoder (i.e., SUD), which exhibits
the lowest decoding complexity for the SM-SCMA system. This algorithm
provides an acceptable BER performance under some practical constraints
(i.e., having good link quality or a high number of receive antennas);
\item Propose an enhanced version of the first algorithm (i.e., MSUD), which
considerably improves the BER performance with a little increase in
the decoding complexity;
\item Form the tree-search decoder for the SM-SCMA system, which is very
important for the SD algorithms that can be investigated in the future
by researchers;
\item Propose an SD algorithm based on the tree-search concept (i.e., FCSD),
which provides a near-optimum BER performance with a significant reduction
in the decoding complexity;
\item Provide the mathematical formulation, pseudo-codes, and complexity
analysis for all these proposed algorithms;
\item Provide Monte Carlo simulation results to indicate the significant
benefits of the proposed algorithms.
\end{enumerate}
The rest of the paper\footnote{Notations: Boldface lowercase and uppercase letters represent vectors
and matrices, respectively. $\mathcal{C}\mathcal{N}$ denotes a complex-valued
normal random variable. $\mathrm{\text{diag}}(\cdotp)$ converts a
vector into a diagonal matrix with diagonal elements that are the
same as the original vector elements. $\left\Vert \centerdot\right\Vert $
denotes the Euclidean norm. $\text{card}\left\{ \centerdot\right\} $
is the is the cardinality of a set that refers to the number of elements
in that set. $[\centerdot]^{\text{T}}$ denotes the matrix or vector
transpose. $\mathbb{E}\left\{ \centerdot\right\} $ denotes the expectation
operation. $\mathcal{P}(\centerdot)$ is the probability of an event.
$f(\centerdot)$ denotes the probability density function (pdf) of
a random variable. $\phi$ is the empty set.} is organized as follows: In Section \ref{sec:System-Model}, the
system model of the uplink SM-SCMA transmitter and receiver is summarized.
In Section \ref{sec:The-Proposed-Decoding}, the proposed decoding
algorithms for the SM-SCMA system are introduced. In Section \ref{sec:Complexity-Analysis},
the complexity analysis of the proposed decoding algorithms are derived
in terms of the number of real additions and multiplications. The
simulation results and conclusions are provided in Sections \ref{sec:Simulation-Results}
and \ref{sec:Conclusions}, respectively.

\section{\label{sec:System-Model}System Model}

In this section, the transmitter and receiver of the uplink SM-SCMA
system are discussed. Assume that $U$   users are sharing $R$ OREs,
where $U>R$. Each of these   users has an unparalleled multidimensional
codebook, $\mathbf{C}^{u}\in\mathbb{C}^{R\times M}$, $u=1,\ldots,U$,
with $\mathbf{c}_{m}^{u}\in\mathbb{C}^{R\times1}$, $m=1,\ldots,M$
as codewords within the codebook and $M$ as the number of codewords.
Since $\mathbf{c}_{m}^{u}$ is sparse, the number of non-zero elements
for each codeword is denoted by $d_{v}$, whereas the number of zero
elements is $R-d_{v}$. It should be noted that the positions of zero
and non-zero elements are fixed for a codebook (i.e., for a user),
and vary from codebook to another to provide a fixed number of overlapped
users per ORE of $\forall R$. In this paper, the number of overlapped
users per ORE is denoted by $d_{f}$.

\subsection{Transmitted and Received Signal}

Fig. \ref{fig:MPA-factor-graph} shows the block diagram of the uplink
SM-SCMA system with $U$ users. Consider an $N_{r}\times N_{t}$ MIMO
system for each user, where $N_{t}$ and $N_{r}$ represent the number
of transmit and receive antennas, respectively. For the $u$-th user
in the SM-SCMA transmitter, the first $\log_{2}(N_{t})$ of the input
bits select the transmit antenna to be activated, while the remaining
$\log_{2}(M)$ bits are mapped to choose a corresponding codebook,
$\mathbf{c}_{m}^{u}$, to be transmitted from that active antenna.
Hence, the spectral efficiency of the $u$-th user is given by

\begin{equation}
\eta_{u}=\log_{2}(N_{t})+\log_{2}(M),\label{eq:SE}
\end{equation}

\noindent where $\eta_{u}$ is the spectral efficiency of the $u$-th
user that is measured in bit per channel use (bpcu). It should be
noted that the total system spectral efficiency for all users is $U\eta_{u}$
bpcu.

At the receiver, the noisy received signal at the $n_{r}$-th receive
antenna of the $r$-th ORE, $y_{n_{r}}^{r}$, is

\begin{equation}
y_{n_{r}}^{r}=\sum_{u\in\varLambda_{r}}\left(h_{n_{r},n_{t}^{u}}^{r,u}c_{m}^{r,u}\right)+n_{n_{r}}^{r},\,\,\,\,\,\,\,r=1,\ldots,R,\label{eq: y_n_r}
\end{equation}

\noindent where $h_{n_{r},n_{t}^{u}}^{r,u}$ represents the Rayleigh
fading channel coefficient between the $n_{r}\text{-th}\in\{1,\ldots,N_{r}\}$
receive antenna and $n_{t}^{u}\text{-th}\in\{1,\ldots,N_{t}\}$ transmit
antenna of the $u$-th user for the $r$-th ORE, $c_{m}^{r,u}$ is
the non-zero $r$-th element for the $m$-th codeword of the $u$-th
user. Here, $\varLambda_{r}$ denotes the set of users indices that
share the $r$-th ORE, and $n_{n_{r}}^{r}\sim\mathcal{C}\mathcal{N}\left(0,\sigma^{2}\right)$
is the complex additive white Gaussian noise (AWGN) with zero-mean
and a variance of $\sigma^{2}$ for the $r$-th ORE at the $n_{r}$-th
receive antenna.

For all OREs, the received signal at the $n_{r}$-th receive antenna,
$\mathbf{y}_{n_{r}}\in\mathbb{C}^{R\times1}=[y_{n_{r}}^{1},\ldots,y_{n_{r}}^{R}]^{\text{T}}$,
is given by

\begin{equation}
\mathbf{y}_{n_{r}}=\sum_{u=1}^{U}\left(\text{diag}\left(\mathbf{h}_{n_{r},n_{t}^{u}}^{u}\right)\mathbf{c}_{m}^{u}\right)+\mathbf{n}_{n_{r}},\label{eq: y_n}
\end{equation}

\noindent where $\mathbf{h}_{n_{r},n_{t}^{u}}^{u}\in\mathbb{C}^{R\times1}=[h_{n_{r},n_{t}^{u}}^{1,u},\ldots,h_{n_{r},n_{t}^{u}}^{R,u}]^{\text{T}}$
is the Rayleigh fading channel vector between the $n_{r}\text{-th}$
receive antenna and $n_{t}^{u}\text{-th}$ transmit antenna of the
$u$-th user, and $\mathbf{n}_{n_{r}}\in\mathbb{C}^{R\times1}=[n_{n_{r}}^{1}\ldots n_{n_{r}}^{R}]^{\text{T}}$
is the AWGN vector.

It is worth noting that the relationship between the position of zero/non-zero
elements of users codebooks and OREs can be described by a binary
indicator matrix, $F$. In the indicator matrix, the number of rows
and columns represents the number of OREs and number of users, respectively.
Moreover, the ones in $F$ show the position of non-zero elements
of the user codebooks. In this paper, six users overloaded over four
OREs (i.e., $U=6$ and $R=4$) are considered, with $F$ given by
{[}\ref{H.-Nikopour-and}{]}, {[}\ref{I.-Al-Nahhal,-O. TVT}{]}:

\begin{equation}
F=\left[\begin{array}{cccccc}
0 & 1 & 1 & 0 & 1 & 0\\
1 & 0 & 1 & 0 & 0 & 1\\
0 & 1 & 0 & 1 & 0 & 1\\
1 & 0 & 0 & 1 & 1 & 0
\end{array}\right].\label{eq: indicator matrix}
\end{equation}

\noindent As seen from (\ref{eq: indicator matrix}), $d_{v}=2$ for
all users and $d_{f}=3$ for all OREs. A useful representation for
the indicator matrix is

\begin{equation}
\varLambda_{r}=\left\{ \varLambda_{r}(1),\ldots,\varLambda_{r}(d_{f})\right\} ,\label{eq: A general}
\end{equation}

\noindent where $\varLambda_{r}(1)$ denotes the index of the first
user that shares the $r$-th ORE, and $\text{card}\{\varLambda_{r}\}=d_{f}$.
Thus, $F$ in (\ref{eq: indicator matrix}) yields

\begin{subequations}
\label{A}
\begin{align}
\varLambda_{1} & =\left\{ \varLambda_{1}(1),\varLambda_{1}(2),\varLambda_{1}(3)\right\} =\left\{ 2,3,5\right\} ,\label{eq: A_1}\\
\varLambda_{2} & =\left\{ \varLambda_{2}(1),\varLambda_{2}(2),\varLambda_{2}(3)\right\} =\left\{ 1,3,6\right\} ,\label{eq: A_2}\\
\varLambda_{3} & =\left\{ \varLambda_{3}(1),\varLambda_{3}(2),\varLambda_{3}(3)\right\} =\left\{ 2,4,6\right\} ,\label{eq: A_3}\\
\varLambda_{4} & =\left\{ \varLambda_{4}(1),\varLambda_{4}(2),\varLambda_{4}(3)\right\} =\left\{ 1,4,5\right\} .\label{eq: A_4}
\end{align}
\end{subequations}

\subsection{Signal Detection}

At the receiver side, the decoder task is to estimate the activated
transmit antenna and the mapped codeword for each user (i.e., user
message). In this subsection, the ML and MPA decoders are discussed.

\subsubsection{ML Decoder}

The ML decoder jointly performs an exhaustive search for all possible
combinations between the transmit antennas and codewords for all users
(i.e., $\left(N_{t}M\right)^{U}$ possible combinations). Although
the ML provides the optimum BER performance, it has an impractically
high decoding complexity. The mathematical formulation of the ML decoder
is given by

$\vphantom{}$

$\left\{ \hat{\mathbf{C}},\hat{\mathbf{j}}\right\} =\underset{\begin{array}{c}
j=1,\ldots,N_{t}^{U}\\
l=1,\ldots,M^{U}
\end{array}}{\text{arg}\,\text{\,min}}$

\begin{equation}
\left\{ \sum_{n_{r}=1}^{N_{r}}\left\Vert \mathbf{y}_{n_{r}}-\sum_{u=1}^{U}\left(\text{diag}\left(\mathbf{h}_{n_{r},n_{t}^{u}(j)}^{u}\right)\mathbf{c}_{m(l)}^{u}\right)\right\Vert ^{2}\right\} ,\label{eq:ML}
\end{equation}

\noindent where $\hat{\mathbf{j}}=\{\hat{n_{t}}^{1},\ldots,\hat{n_{t}}^{U}\}$
denotes the set of indices of the estimated active transmit antenna
for all $U$ users, with $\hat{n_{t}}^{u}$ as the estimated index
of the active transmit antenna for the $u$-th user, $n_{t}^{u}(j)$
is the active transmit antenna index of the $u$-th user that corresponds
to the $j$-th antenna combinations (out of $(N_{t})^{U}$ combinations)
of all $U$ users, $\hat{\mathbf{C}}\in\mathbb{C}^{R\times U}=[\hat{\mathbf{c}}_{m}^{1}\ldots\hat{\mathbf{c}}_{m}^{U}]$
represents the estimated transmitted codewords of the $U$ users,
with $\hat{\mathbf{c}}_{m}^{u}$ as the estimated transmitted codeword
of the $u$-th user, and $m(l)$ is the $m$-th codeword of the $u$-th
user that corresponds to the $l$-th codeword combinations (out of
$(M)^{U}$ combinations) of all $U$ users.

\subsubsection{MPA Decoder}

\begin{figure*}
\begin{centering}
\includegraphics[scale=0.49]{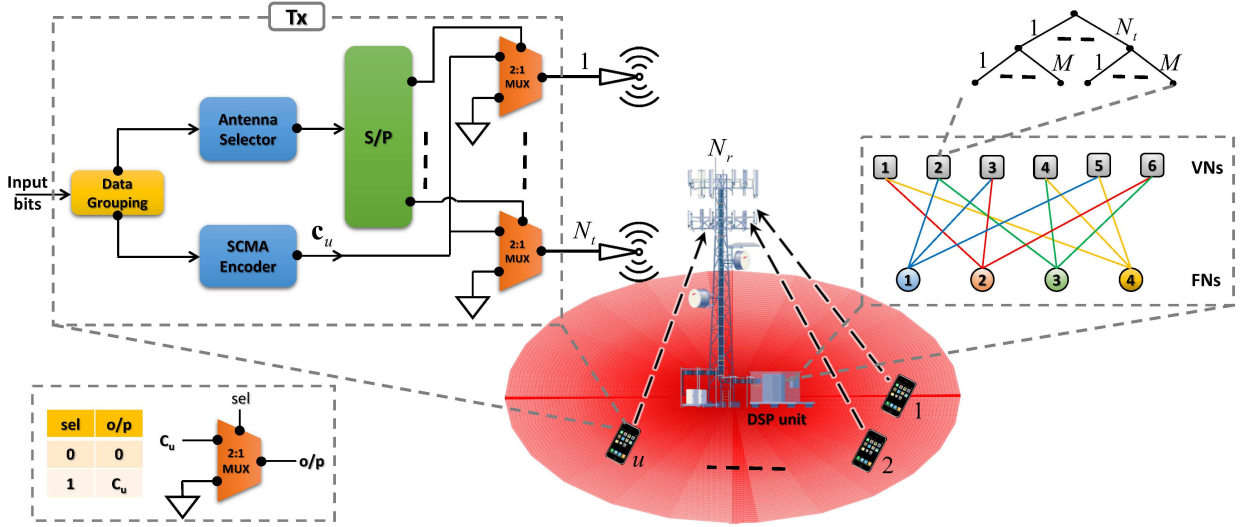}
\par\end{centering}
\caption{\textcolor{blue}{\label{fig:MPA-factor-graph}}Uplink SM-SCMA system.}
\end{figure*}

The MPA is an alternative practical decoder to the ML decoder. It
iteratively updates the probability of users messages between the
function nodes (FNs) that represent the number of OREs, and the variable
nodes (VNs) that represents the number of users. It is worth noting
that each of the FNs is connected with all VNs that share the same
FN based on indicator matrix in (\ref{eq: indicator matrix}) to form
what is called a factor graph. The factor graph of the MPA decoder
used in this paper is shown in Fig. \ref{fig:MPA-factor-graph} for
$U=6$, $R=4$ and $F$, which is given from (\ref{eq: indicator matrix}).

It is assumed that the probability of passing the $u$-th message,
$\{c_{m}^{r,u},n_{t}^{u}\}$, from the $u$-th VN to the $r$-th FN
and vice versa at the $k$-th iteration (out of $K$ iterations) is
$\mathcal{P}_{v_{u}\rightarrow f_{r}}^{(k)}(\{c_{m}^{r,u},n_{t}^{u}\})$
and $\mathcal{P}_{f_{r}\rightarrow v_{u}}^{(k)}(\{c_{m}^{r,u},n_{t}^{u}\})$,
respectively. Initially, all users messages passing from the VNs to
FNs are equiprobable, i.e.,

\begin{equation}
\mathcal{P}_{v_{u}\rightarrow f_{r}}^{(0)}\left(\{c_{m}^{r,u},n_{t}^{u}\}\right)=\frac{1}{N_{t}M},\,\,\,\,\,\forall u,\,\,\forall r,\,\,\forall m.\label{eq:Equal_prob}
\end{equation}

\noindent The mathematical formulation of updating the messages at
the $(k+1)$-th iteration of the MPA decoder is given by {[}\ref{Z.-Pan,-J.}{]},
{[}\ref{I.-Al-Nahhal,-O. TVT}{]}:

\[
\mathcal{P}_{f_{r}\rightarrow v_{u}}^{(k+1)}\left(\left\{ c_{m}^{r,u},n_{t}^{u}\right\} \right)=\,\hspace{5cm}
\]

\[
\sum_{\psi(i),i\in\left.\varLambda_{r}\right\backslash u}\left\{ \prod_{n_{r}=1}^{N_{r}}\left(\mathcal{P}\left(\mathbf{y}_{n_{r}}|\psi(i),\psi(u)=\{c_{m}^{r,u},n_{t}^{u}\}\right)\right)\right.
\]

\begin{equation}
\left.\times\prod_{i\in\left.\varLambda_{r}\right\backslash u}\mathcal{P}_{v_{i}\rightarrow f_{r}}^{(k)}\left(\psi(i)\right)\right\} ,\,\,\,\,\,\forall m,\,\,\forall r,\,\,u\in\varLambda_{r},\label{eq:FN_to_VN}
\end{equation}

\noindent where $\left.\varLambda_{r}\right\backslash u$ represents
$\varLambda_{r}$ in (\ref{eq: A general}) except the $u$-th user
and $\psi(\centerdot)$ represents the message of a user. The conditional
probability in (\ref{eq:FN_to_VN}) is given by

\begin{equation}
\mathcal{P}\left(\mathbf{y}_{n_{r}}|\boldsymbol{\psi}^{r}\right)=\frac{1}{\sqrt{2\pi}\sigma}\,\text{exp}\hspace{-0.1cm}\left(\hspace{-0.1cm}-\frac{\left|y_{n_{r}}^{r}-\hspace{-0.1cm}\sum_{u\in\varLambda_{r}}\hspace{-0.1cm}\left(h_{n_{r},n_{t}^{u}}^{r,u}c_{m}^{r,u}\right)\right|^{2}}{2\sigma^{2}}\right),\label{eq:Condtional_Prob}
\end{equation}

\noindent where $\boldsymbol{\psi}^{r}$ represents the possible messages
of all users that share the $r$-th ORE.

Now, $\mathcal{P}_{v_{u}\rightarrow f_{r}}^{(k+1)}(\{c_{m}^{r,u},n_{t}^{u}\})$
can be calculated as

$\vphantom{}$

$\mathcal{P}_{v_{u}\rightarrow f_{r}}^{(k+1)}\left(\left\{ c_{m}^{r,u},n_{t}^{u}\right\} \right)=\gamma_{u,r}^{(k+1)}$

\begin{equation}
\times\prod_{j\in\left.\Omega_{u}\right\backslash r}\hspace{-0.2cm}\mathcal{P}_{f_{r}\rightarrow v_{u}}^{(k+1)}\left(\left\{ c_{m}^{r,u},n_{t}^{u}\right\} \right),\hspace{0.5cm}\forall m,\,\,\forall u,\,\,r\in\Omega_{u},\label{eq:VN_to_FN}
\end{equation}

\noindent where $\Omega_{u}$ denotes the set of ORE indices that
correspond to $d_{v}$ non-zero positions for the $u$-th user, $\left.\Omega_{u}\right\backslash r$
represents the set $\Omega_{u}$ except the $r$-th ORE, and $\gamma_{u,r}^{(k+1)}$
is

\begin{equation}
\gamma_{u,r}^{(k+1)}=\left(\sum_{m=1}^{M}\sum_{n_{t}=1}^{N_{t}}\mathcal{P}_{v_{u}\rightarrow f_{r}}^{(k)}\left(\left\{ c_{m}^{r,u},n_{t}^{u}\right\} \right)\right)^{-1}.\label{eq: normalization factor}
\end{equation}

After the MPA completes $K$ iterations, the estimated message of
the $u$-th user can be calculated by

\begin{equation}
\left\{ \hat{\mathbf{c}}_{m}^{u},\hat{n_{t}}^{u}\right\} ^{(K)}=\hspace{-0.4cm}\hspace{-0.2cm}\underset{\begin{array}{c}
m=1,\ldots,M\\
n_{t}=1,\ldots,N_{t}
\end{array}}{\text{arg}\,\text{max}}\hspace{-0.2cm}\prod_{j\in\Omega_{u}}\mathcal{P}_{f_{j}\rightarrow v_{u}}^{(K)}\left(\left\{ c_{m}^{r,u},n_{t}^{u}\right\} \right),\,\,\,\forall u.\label{eq: Final_MPA}
\end{equation}

\noindent The set of all estimated users messages using the MPA in
(\ref{eq: Final_MPA}), $\hat{\Theta}_{\text{MPA}}$, can be given
as

\begin{equation}
\hat{\Theta}_{\text{MPA}}=\left\{ \left\{ \hat{\mathbf{c}}_{m}^{1},\hat{n_{t}}^{1}\right\} ^{(K)},\,\ldots,\,\{\hat{\mathbf{c}}_{m}^{U},\hat{n_{t}}^{U}\}^{(K)}\right\} .\label{eq: Theta MPA}
\end{equation}

\section{\label{sec:The-Proposed-Decoding}The Proposed Decoding Algorithms}

In this section, the three proposed decoding algorithms are introduced.
The first two algorithms focus on decoding the signal with very low
complexity and acceptable BER performance. The third proposed algorithm
employs the SD concept to provide a near-optimum BER performance with
low-decoding complexity in addition to other advantages, such as the
feasibility of parallel hardware implementation and the flexible trade-off
between decoding complexity and BER performance.

\subsection{The SUD Algorithm}

The SUD algorithm provides the lowest decoding complexity among the
proposed algorithms at the expense of BER performance. It successively
detects the users' messages using only one ORE. Then, the SUD algorithm
uses these detected messages as given information in the next OREs
to detect the rest of the users' messages. Consequently, the SUD algorithm
does not benefit from the diversity gain (i.e., sharing the information
over several OREs) of the users' codebook. For instance, if a user
spreads his message over $d_{v}$ OREs (as in (\ref{eq: indicator matrix})),
the SUD algorithm uses only one ORE to detect this message. Thus,
the SUD algorithm considers that the user's message is given for the
rest of the shared OREs (i.e., $d_{v}-1$ OREs).

At the beginning, the SUD algorithm performs an exhaustive search
for all combinations of the users messages that share the first ORE.
It starts with the OREs with highest energy, $E^{r}$, based on the
following

\begin{equation}
E^{r}=\sum_{u\in\varLambda_{r}}\sum_{n_{r}=1}^{N_{r}}\left|h_{n_{r},n_{t}^{u}}^{r,u}\right|^{2},\hspace{0.7cm}r=1,\ldots,R.\label{eq: energy}
\end{equation}

\noindent Then, these estimated users messages are employed to estimate
the messages of other users that share the next OREs. Sequentially,
the SUD algorithm estimates the undetected users messages until they
are all estimated based on the descending order of $E^{r}$ in (\ref{eq: energy})
for $\forall R$.

The mathematical formulation of the SUD algorithm is given by

$\vphantom{}$

$\left\{ \hat{\mathbf{C}}^{r},\hat{\mathbf{j}}^{r}\right\} =\underset{\begin{array}{c}
j=1,\ldots,N_{t}^{\grave{U}^{r}}\\
l=1,\ldots,M^{\grave{U}^{r}}
\end{array}}{\text{arg}\,\text{\,min}}$

\noindent 
\[
\left\{ \sum_{n_{r}=1}^{N_{r}}\left|y_{n_{r}}^{r}-\underset{\text{Term 1}}{\underbrace{\sum_{u\in\grave{\varLambda}_{r}}h_{n_{r},\hat{n}_{t}^{u}}^{r,u}c_{\hat{m}}^{r,u}}}-\underset{\text{Term 2}}{\underbrace{\sum_{u\in\varLambda_{r}\backslash\grave{\varLambda}_{r}}h_{n_{r},n_{t}^{u}(j)}^{r,u}c_{m(l)}^{r,u}}}\right|^{2}\right\} {\normalcolor },
\]

\begin{equation}
\hspace{4.5cm}1\leq r\leq R,\label{eq: SUD}
\end{equation}

\noindent where $\grave{\varLambda}_{r}$ is the set of users indices
that share the $r$-th ORE in which their messages are already estimated
previously, $\varLambda_{r}\backslash\grave{\varLambda}_{r}$ is $\varLambda_{r}$
except $\grave{\varLambda}_{r}$ or it is the set of users indices
that share the $r$-th ORE and their messages need to be estimated,
$\grave{U}^{r}=\text{card}\{\varLambda_{r}\backslash\grave{\varLambda}_{r}\}\leq d_{f}$
is the number of users whose messages need to be estimated at the
$r$-th ORE, $\hat{\mathbf{j}}^{r}$ represents the set of indices
of the estimated active transmit antennas for all $\grave{U}^{r}$
users at the $r$-th ORE, and $\hat{\mathbf{C}}^{r}$ denotes the
estimated transmitted codewords of the $\grave{U}^{r}$ users at the
$r$-th ORE. Here, $\text{Term 1}$ and $\text{Term 2}$ represent
the users messages that have already been estimated from previous
OREs and that need to be estimated at the $r$-th ORE, respectively.
It is worth noting that $\text{Term 1}$ equals zero at the first
ORE used by the SUD algorithm (i.e., $\grave{\varLambda}_{1}=\phi$).
After estimating all users messages from certain OREs, the set of
complete estimated users messages using the SUD algorithm in (\ref{eq: SUD}),
$\hat{\Theta}_{\text{SUD}}$, can be written as

\begin{equation}
\hat{\Theta}_{\text{SUD}}=\left\{ \left\{ \hat{\mathbf{c}}_{m}^{1},\hat{n_{t}}^{1}\right\} ,\,\ldots,\,\{\hat{\mathbf{c}}_{m}^{U},\hat{n_{t}}^{U}\}\right\} .\label{eq: Theta SUD}
\end{equation}

Consequently, the SUD algorithm detects users messages using a single
ORE. Then, these detected messages are used as given messages to detect
the others that share the rest of the $d_{v}-1$ OREs. It should be
noted that the SUD algorithm may not use all received signals on OREs
if all users messages are already estimated using certain OREs. The
SUD algorithm is summarized in Algorithm \ref{alg: SUD}.

\begin{algorithm}[t]
\begin{itemize}
\item \textbf{Store} codebooks for all users;
\item \textbf{Input} channel matrices for all users;
\item \textbf{Define}\textbf{\small{} }{\small{}$\hat{\Theta}_{\text{SUD}}$}
and \textbf{$\varLambda$} as the set of estimated users messages
and set of users indices corresponding to the estimated messages in
$\hat{\Theta}_{\text{SUD}}$, respectively;
\item \textbf{Initialize }$\hat{\Theta}_{\text{SUD}}=\{\cdot\}$ and $\varLambda=\{\cdot\}$;
\item \textbf{Order} the OREs which should be visited based on (\ref{eq: energy});
\end{itemize}
~~~~~1: \textbf{While$\,\,\,r\leq R$, do}

~~~~~2:\textbf{ ~~~Set} $\grave{\varLambda}_{r}\leftarrow\{\varLambda\cap\varLambda_{r}\}$;

~~~~~3: ~~~\textbf{Assign }$\bar{y}_{n_{r}}^{r}\leftarrow y_{n_{r}}^{r}-\sum_{u\in\grave{\varLambda}_{r}}h_{n_{r},\hat{n}_{t}^{u}}^{r,u}c_{\hat{m}}^{r,u}$;

~~~~~4:\textbf{ ~~~Find }$\{\hat{\mathbf{C}}^{r},\hat{\mathbf{j}}^{r}\}$
that solves the following:

~~~~~~~\textbf{ ~~~}$\underset{j\,\&\,l}{\text{arg}\,\text{min}}\{\sum_{n_{r}=1}^{N_{r}}|\bar{y}_{n_{r}}^{r}-\sum_{u\in\varLambda_{r}\backslash\grave{\varLambda}_{r}}h_{n_{r},n_{t}^{u}(j)}^{r,u}c_{m(l)}^{r,u}|^{2}\}$

~~~~~~~\textbf{ ~~}s.t. $j=1,\ldots,N_{t}^{\grave{U}^{r}}$
and $l=1,\ldots,M^{\grave{U}^{r}}$;

~~~~~5:\textbf{ ~~~Update} $\hat{\Theta}_{\text{SUD}}$ based
on $\{\hat{\mathbf{C}}^{r},\hat{\mathbf{j}}^{r}\}$;

~~~~~6:\textbf{ ~~~Update} \textbf{$\varLambda$ }based on\textbf{
}$\hat{\Theta}_{\text{SUD}}$;

~~~~~7: ~~~\textbf{ if} $\text{card}\{\varLambda\}==U$

~~~~~8:\textbf{ ~~~~~ break }and end the algorithm;

~~~~~9: ~~~\textbf{ end if}

~~~~10: \textbf{~~ Set $r\leftarrow r+1$;}

~~~~11: \textbf{end While}
\begin{itemize}
\item \textbf{Output} $\hat{\Theta}_{\text{SUD}}$.
\end{itemize}
\caption{\label{alg: SUD}The proposed SUD algorithm pseudo-code.}
\end{algorithm}

\subsection{The MSUD Algorithm}

As mentioned in the SUD algorithm, the user's message is detected
using a single ORE; however, the $(d_{v}-1)$ non-zero OREs for each
user are not included in the decoding process with the aim of reducing
the decoding complexity. This leads to a significant deterioration
in the BER performance (i.e., losing the diversity gain). Unlike the
SUD algorithm, the MSUD algorithm considers all OREs that carry the
user's message for detection, to improve the BER performance.

The MSUD is an iterative algorithm that estimates the user message
by considering only one user message unknown at a time. In contrast,
the rest of the users messages are considered to be known from the
previous iteration. Moreover, the MSUD algorithm considers the received
signals from all $d_{v}$ non-zero OREs for each user in the detection
process to improve the BER performance. It is important to mention
that the initial values of the users messages used in the MSUD algorithm
are estimated using the SUD algorithm. In other words, the MSUD algorithm
performs the SUD algorithm first. Then, $K$ iterations are performed
to improve the BER performance.

To formulate the MSUD algorithm, user messages are first estimated
from the SUD algorithm (i.e., $\hat{\Theta}_{\text{SUD}}$, in (\ref{eq: Theta SUD}))
and are subsequently used as input to/initialization of the iteration
stage of the MSUD algorithm. At the $(k+1)$-th iteration, the estimated
$u$-th user message, $\{\hat{\mathbf{c}}_{m}^{u},\hat{n_{t}}^{u}\}^{(k+1)}$,
is given by

$\vphantom{}$

$\left\{ \hat{\mathbf{c}}_{m}^{u},\hat{n_{t}}^{u}\right\} ^{(k+1)}=\underset{\begin{array}{c}
j=1,\ldots,N_{t}\\
l=1,\ldots,M
\end{array}}{\text{arg}\,\text{\,min}}$

\[
\left\{ \sum_{r\in\Omega_{u}}\sum_{n_{r}=1}^{N_{r}}\left|y_{n_{r}}^{r}-\hspace{-3mm}\underset{\text{Term 3}}{\underbrace{\sum_{\grave{u}\in\varLambda_{r}\backslash u}\left\{ h_{n_{r},\hat{n}_{t}^{\grave{u}}}^{r,\grave{u}}c_{\hat{m}}^{r,\grave{u}}\right\} ^{\left(k\right)}}}\hspace{-1.5mm}-\underset{\text{Term 4}}{\underbrace{h_{n_{r},n_{t}^{u}(j)}^{r,u}c_{{\color{blue}{\color{blue}{\color{black}m(l)}}}}^{r,u}}}\right|^{2}\right\} ,
\]

\begin{equation}
\hspace{5.5cm}u=1,\,\ldots,\,U,\label{eq: MSUD}
\end{equation}

\noindent where $\text{Term 3}$ and $\text{Term 4}$ represent the
given estimated users messages that share the same ORE with the $u$-th
user and the desired user message of the $u$-th user to be estimated,
respectively. The MSUD algorithm uses all $d_{v}$ non-zero OREs for
each user in the detection, which can be seen from $\sum_{r\in\Omega_{u}}$
in (\ref{eq: MSUD}). The estimation process using (\ref{eq: MSUD})
is performed for all $U$ users for each iteration. After $K$ iterations,
the set of estimated messages for all $U$ users, $\hat{\Theta}_{\text{MSUD}}$,
is

\begin{equation}
\hat{\Theta}_{\text{MSUD}}=\left\{ \left\{ \hat{\mathbf{c}}_{m}^{1},\hat{n_{t}}^{1}\right\} ^{(K)},\,\ldots,\,\{\hat{\mathbf{c}}_{m}^{U},\hat{n_{t}}^{U}\}^{(K)}\right\} .\label{eq: Theta MSUD}
\end{equation}

For example, assume that we need to detect the message of the second
user (i.e., $u=2$) using the MSUD algorithm. From (\ref{eq: A_1})
and (\ref{eq: A_3}), the received signal of the first and third OREs
will be considered in the detection, while the rest of the overlapped
users' messages (i.e., $u=3$, 5, 4 and 6) will be given from the
previous iteration. Algorithm \ref{alg: MSUD} shows the summary of
the MSUD algorithm.

\subsection{The FCSD Algorithm}

The MPA decoder has a limited support to the parallel hardware implementation,
where all users messages are detected together after iterative sequential
stages, as seen from (\ref{eq:FN_to_VN}), (\ref{eq:VN_to_FN}) and
(\ref{eq: Final_MPA}). In practice, this kind of hardware implementation
is not preferable. Besides, the MPA decoder provides a limited trade-off
between decoding complexity and BER performance, which limits its
practicality for applications with specific requirements.

The FCSD algorithm supports the parallel hardware implementation and
also provides a flexible trade-off between decoding complexity and
BER performance. To clearly understand the concept of the FCSD algorithm,
a tree-search for the SM-SCMA should be constructed first.

\begin{algorithm}[t]
\begin{itemize}
\item \textbf{Store} codebooks for all users;
\item \textbf{Input} channel matrices for all users;
\item \textbf{Perform} Algorithm \ref{alg: SUD} to obtain $\hat{\Theta}_{\text{SUD}}$;
\item \textbf{Initialize }$\hat{\Theta}_{\text{MSUD}}=\hat{\Theta}_{\text{SUD}}$;
\end{itemize}
{\small{}~~~~~1:}\textbf{\textit{\small{} }}\textbf{\textit{For
$k=1:K$, do}}

{\small{}~~~~~2:~~~ }\textbf{\textit{For$\,\,\,u=1:U$, do}}

{\small{}~~~~~3: ~~~~~~}\textbf{ Assign }$\bar{y}_{n_{r}}^{r}\leftarrow y_{n_{r}}^{r}-\sum_{\grave{u}\in\varLambda_{r}\backslash u}\left\{ h_{n_{r},\hat{n}_{t}^{\grave{u}}}^{r,\grave{u}}c_{\hat{m}}^{r,\grave{u}}\right\} ^{k}$;

{\small{}~~~~~4:}\textbf{\small{} ~~~~~~}\textbf{Find}\textbf{\small{}
}$\left\{ \hat{\mathbf{c}}_{m}^{u},\hat{n_{t}}^{u}\right\} ^{(k)}${\small{}
}that solves the following:

{\small{}~~~~~~~}\textbf{\small{} ~~~}~~~$\underset{j\,\&\,l}{\text{arg}\,\text{min}}\{\sum_{r\in\Omega_{u}}\sum_{n_{r}=1}^{N_{r}}|\bar{y}_{n_{r}}^{r}-h_{n_{r},n_{t}^{u}(j)}^{r,u}c_{m(l)}^{r,u}|^{2}\}$

{\small{}~~~~~~~}\textbf{\small{} ~~~~~~}s.t. $j=1,\ldots,N_{t}$
and $l=1,\ldots,M$;

{\small{}~~~~~5:}\textbf{\small{} ~~~~~~}\textbf{Update}
$\hat{\Theta}_{\text{MSUD}}$ based on $\left\{ \hat{\mathbf{c}}_{m}^{u},\hat{n_{t}}^{u}\right\} ^{(k)}$;

{\small{}~~~~~6:~~~ }\textbf{\textit{end For}}

{\small{}~~~~~7: }\textbf{\textit{end For}}
\begin{itemize}
\item \textbf{Output}{\small{} }$\hat{\Theta}_{\text{MSUD}}$.
\end{itemize}
\caption{\label{alg: MSUD}The proposed MSUD algorithm pseudo-code.}
\end{algorithm}

\subsubsection{SM-SCMA Tree-search}

\begin{figure}
\begin{centering}
\includegraphics[scale=0.27]{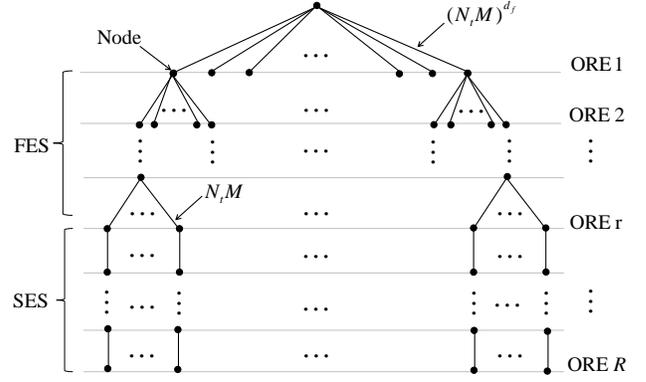}
\par\end{centering}
\caption{\label{fig:Tree-search-of-the}The proposed tree-search for the SM-SCMA
system.}
\end{figure}

The ML decoder of the SM-SCMA in (\ref{eq:ML}) can be represented
as a multi-level tree-search, as in Fig. \ref{fig:Tree-search-of-the}.
Each of the tree-search levels corresponds to an ORE (i.e., the number
of levels equals $R$). At each level, there is a certain number of
nodes representing the distance metric between the received signal
at the $r$-th ORE and possible combinations of the users messages
that share this ORE. Each node at the $r$-th level is expanded into
child nodes at the next level.

The mathematical formulation of the $i$-th node at the $r$-th level,
$d_{i}^{r}$, is

\begin{equation}
d_{i}^{r}=d_{i}^{r-1}+e_{i}^{r},\,\,\,\,\,\,r=1,\,\ldots,\,R,\label{eq: Acumalated Nod}
\end{equation}

\noindent where $d_{i}^{r-1}$ is the mother node of $d_{i}^{r}$
and $e_{i}^{r}$ is given by

\begin{equation}
e_{i}^{r}=\sum_{n_{r}=1}^{N_{r}}\left|y_{n_{r}}^{r}-\hspace{-1mm}\sum_{u\in\grave{\varLambda}_{r}}h_{n_{r},\hat{n}_{t}^{u}}^{r,u}c_{\hat{m}}^{r,u}-\hspace{-3mm}\sum_{u\in\varLambda_{r}\backslash\grave{\varLambda}_{r}}h_{n_{r},n_{t}^{u}(i)}^{r,u}c_{m(i)}^{r,u}\right|^{2}.\label{eq: ED}
\end{equation}

\noindent At the first level (i.e., $r=1$), $d_{i}^{0}=0$ in calculating
$d_{i}^{1}$ and $i=1,\,\ldots,\,(MN_{t})^{d_{f}}$. From (\ref{eq: Acumalated Nod})
and (\ref{eq: ED}), it should be noted that a node is an accumulation
of the distance metric of all preceding nodes in the same branch and
that the value of $e_{i}^{r}$ increases as $r$ increases, respectively.

Unlike the construction of the SM tree-search {[}\ref{I.-Al-Nahhal,-E. JSAC}{]}
and MIMO tree-search {[}\ref{I.-Al-Nahhal,-A. comex}{]}-{[}\ref{I.-Al-Nahhal,-M. Kbest ill condition}{]},
the number of expanded nodes for each mother node of the SM-SCMA tree-search,
as seen in Fig. \ref{fig:Tree-search-of-the} gradually reduces across
the OREs, reaching a limit of one. Consequently, the SM-SCMA tree-search
consists of two stages. The upper stage in which each of the mother
nodes is fully expanded to multiple child nodes, is referred to as
fully expanded stage (FES). The lower stage is called single expanded
stage (SES), in which each of the mother nodes is expanded to only
one node. Typically, each level of FES has at least one or more users
messages that have not been estimated from the previous OREs; the
number of these users messages gradually decreases as the ORE increases,

\begin{equation}
d_{f}\geq\grave{U}^{2}\geq\cdots\geq\grave{U}^{r}>1,\,\,\,\,\,\,\,r\in\text{FES}.\label{eq: Card FES}
\end{equation}

\noindent It is worth noting that the number of nodes at the first
level is $(N_{t}M)^{d_{f}}$ since there are $d_{f}$ users sharing
ORE 1 and no users messages have been estimated previously.

\subsubsection{The FCSD Algorithm}

In the tree-search provided in Fig. \ref{fig:Tree-search-of-the},
the ML solution in (\ref{eq:ML}) (close to the MPA solution) can
be achieved by visiting all nodes, which is extremely high in terms
of decoding complexity. The basic concept of the FCSD algorithm is
to reduce the decoding complexity of the SM-SCMA system by reducing
the search space inside the tree-search based on a predetermined pruned
radius (i.e., threshold). For that, at each level, the nodes that
have values smaller than a certain threshold (i.e., pruned radius)
are the only ones which are expanded at the next level. It is worth
noting that the tree-search levels of Fig. \ref{fig:Tree-search-of-the}
can be ordered based on (\ref{eq: energy}) before performing the
FCSD algorithm.

Let us consider that the pruned radius is denoted by $\boldsymbol{\gamma}\in\mathbb{R}^{R-1}=[\gamma_{1}\ldots\gamma_{r}\ldots\gamma_{R-1}]$
and keeps $[\rho_{1}\ldots\rho_{r}\ldots\rho_{R-1}]$ survived nodes,
where $\gamma_{r}$ is the pruned radius and $\rho_{r}$ is the number
of survived nodes at the $r$-th level. At the final level (i.e.,
the $R$-th level), the minimum node is chosen to be the solution
of the algorithm. Consequently, $\rho_{r}$ for the upper $R-1$ levels
is given by

$\vphantom{}$

\noindent 
\[
\rho_{r}=\left\{ d_{i}^{r}\leq\gamma_{r}|i=1,\ldots,\rho_{r-1}(N_{t}M)^{\grave{U}^{r}}\right\} ,\hspace{3cm}
\]

\begin{equation}
\hspace{3cm}0\leq\grave{U}^{r}\leq d_{f},\,\,\,\,1\leq r\leq R-1,
\end{equation}

\noindent where $\grave{U}^{r}=0$ at $r\in\text{SES}$, $0<\grave{U}^{r}\leq d_{f}$
at $r\in\text{FES}$, and $\rho_{0}=1$ at the first ORE (i.e., $r=1$).
At the last level (i.e., $r=R$), the number of nodes is $\rho_{R-1}$,
since there are only $\rho_{R-1}$ survived nodes from the $R-1$-th
level. Thus, the FCSD algorithm declares the argument of the minimum
node at the last level as the solution, which can be represented as

\begin{equation}
\left\{ \hat{\mathbf{C}},\hat{\mathbf{j}}\right\} =\underset{\begin{array}{c}
i=1,\ldots,\rho_{R-1}\end{array}}{\text{arg}\,\text{\,min}}\left\{ d_{i}^{R}\right\} .\label{eq: FCSD}
\end{equation}

It is worth noting that a higher value of the pruned radius may lead
to expanding unnecessary nodes, which increases the decoding complexity.
On the other hand, a smaller value of the pruned radius may cause
an early dropping of the optimum solution, which deteriorates the
BER performance. Thus, the appropriate choice of the pruned radius
is a crucial process in the FCSD algorithm. For more clarifications,
the accumulated node, $d_{i}^{r}$, in (\ref{eq: Acumalated Nod})
is a non-central chi-squared random variable with $2rN_{r}$ degrees
of freedom and its pdf is given by {[}\ref{J.-Proakis,-Digital},
(Ch. 2){]}

\[
f_{d_{i}^{r}}(d_{i}^{r})=\frac{1}{\sigma^{2}}\left(\frac{d_{i}^{r}}{\alpha_{r,i}^{2}}\right)^{(rN_{r}-1)/2}\hspace{4.3cm}
\]

\begin{equation}
\times\,\text{exp}\left(-\frac{\alpha_{r,i}^{2}+d_{i}^{r}}{\sigma^{2}}\right)\,I_{rN_{r}-1}\left(\frac{\sqrt{d_{i}^{r}\,\alpha_{r,i}^{2}}}{\sigma_{n}^{2}/2}\right),\label{eq: d_i_j pdf}
\end{equation}

\noindent where $I_{rN_{r}-1}\left(\centerdot\right)$ is the first
kind modified Bessel function with order $(rN_{r}-1)$ and the non-centrality
parameter $\alpha_{r,i}^{2}$ is

\[
\alpha_{r,i}^{2}=\sum_{n_{r}=1}^{N_{r}}\sum_{\bar{r}=1}^{r}\left|\sum_{u\in\varLambda_{\bar{r}}}\left(h_{n_{r},n_{t}^{u}}^{\bar{r},u}c_{m}^{\bar{r},u}\right)\right.\hspace{4.3cm}
\]

\begin{equation}
\left.-\hspace{-1mm}\sum_{u\in\grave{\varLambda}_{\bar{r}}}h_{n_{r},\hat{n}_{t}^{u}}^{\bar{r},u}c_{\hat{m}}^{\bar{r},u}-\hspace{-3mm}\sum_{u\in\varLambda_{\bar{r}}\backslash\grave{\varLambda}_{\bar{r}}}h_{n_{r},n_{t}^{u}(i)}^{\bar{r},u}c_{m(i)}^{\bar{r},u}\right|^{2}.\label{eq: non-cenral_parameter}
\end{equation}

\noindent Since $d_{i}^{r}$ has an even degrees of freedom value,
the probability of not dropping the optimum solution early, $d_{i}^{r}|_{\text{opt}}$,
can be calculated as {[}\ref{J.-Proakis,-Digital}, (Ch. 2){]}

\begin{equation}
\mathcal{P}\left(d_{i}^{r}|_{\text{opt}}\leq\gamma_{r}\right)=1-Q_{rN_{r}}\left(\frac{\alpha_{r,i}}{\sigma/\sqrt{2}}\,,\,\frac{\sqrt{\gamma_{r}}}{\sigma/\sqrt{2}}\right),\label{eq: CDF}
\end{equation}

\noindent where $Q_{rN_{r}}(\centerdot,\centerdot)$ is the generalized
Marcum function of order $rN_{r}$. As seen from (\ref{eq: CDF}),
by increasing the value of $\gamma_{r}$, the value of $\mathcal{P}(d_{i}^{r}|_{\text{opt}}\leq\gamma_{r})$
becomes closer to unity.

In the FCSD algorithm, the value of $\gamma_{r}$ is empirically selected
to choose a fixed number of nodes from each level to increase the
probability of including the optimal solution based on (\ref{eq: CDF}).
Accordingly, at each level, the value of $\rho_{r}$ in the FCSD algorithm
is fixed for $1\leq r\leq R-1$. Finally, the FCSD algorithm selects
the minimum node among all expanded nodes at the last level to be
declared as a solution. Thus, the set of estimated messages for all
$U$ users, $\hat{\Theta}_{\text{FCSD}}$, is

\begin{equation}
\hat{\Theta}_{\text{FCSD}}=\left\{ \left\{ \hat{\mathbf{c}}_{m}^{1},\hat{n_{t}}^{1}\right\} ^{(R)},\,\ldots,\,\{\hat{\mathbf{c}}_{m}^{U},\hat{n_{t}}^{U}\}^{(R)}\right\} ,\label{eq: Theta FCSD}
\end{equation}

\noindent where $\{\hat{\mathbf{c}}_{m}^{u},\hat{n_{t}}^{u}\}^{(R)}$
is the estimated message of the $u$-th user corresponding to the
minimum node at the $R$-th level. Algorithm \ref{alg: FCSD} summarizes
the procedure of the FCSD algorithm.

\begin{algorithm}[t]
\begin{itemize}
\item \textbf{Store} codebooks for all users.
\item \textbf{Input} channel matrices for all users.
\item \textbf{Input} $\boldsymbol{\rho}=[\rho_{1}\ldots\rho_{r}\ldots\rho_{R-1}]\in\mathbb{R}^{R-1}$;
\item \textbf{Order} the OREs which should be visited based on (\ref{eq: energy});
\item \textbf{Assign }$\nabla^{r}$ as an empty vector that contains the
distance metric nodes at the $r$-th level;
\item \textbf{Define} $\ell^{r}$ as the total number of nodes in the $r$-th
level;
\end{itemize}
{\small{}~~~~~1: }\textbf{\textit{While$\,\,\,r\leq R-1$, do}}

{\small{}~~~~~2:~~~ }\textbf{\textit{For $i=1:\ell^{r}$,
do}}

{\small{}~~~~~3: ~~~~~~}\textbf{ Compute }$d_{i}^{r}$
from (\ref{eq: Acumalated Nod}) and (\ref{eq: ED});

{\small{}~~~~~4: ~~~~~~}\textbf{ Store}\textbf{\small{}
}{\small{}$d_{i}^{r}$} in $\nabla^{r}$;

{\small{}~~~~~5:~~~ }\textbf{\textit{end For}}

{\small{}~~~~~6:~~~ }\textbf{Keep} the smallest $\rho_{r}$
nodes from $\nabla^{r}$;

{\small{}~~~~~7:~~~ }\textbf{Expand} the survived $\rho_{r}$
nodes from \textbf{\textit{Line \#6}}

{\small{}~~~~~~~~~~ }into $\nabla^{r+1}$;

{\small{}~~~~~8:~~~ }\textbf{Set $r\leftarrow r+1$;}

{\small{}~~~~~9: }\textbf{\textit{end While}}

{\small{}~~~~10: }\textbf{Find} the minimum node in $\nabla^{R}$;
\begin{itemize}
\item \textbf{Output}{\small{} }$\hat{\Theta}_{\text{FCSD}}$ as the messages
corresponding to the argument of the minimum node in \textbf{\textit{Line
\#10}}.
\end{itemize}
\caption{\label{alg: FCSD}The proposed FCSD algorithm pseudo-code.}
\end{algorithm}

\section{\label{sec:Complexity-Analysis}Complexity Analysis}

In this section, the decoding complexities of the conventional MPA
and the proposed algorithms for the SM-SCMA system are discussed.
In this paper, the decoding complexity is measured by the number of
real additions and multiplications required to perform a particular
algorithm. For the conventional MPA decoder of the SM-SCMA system,
the required number of real additions and multiplications, $\text{Add}^{(\text{MPA})}$
and $\text{Mul}^{(\text{MPA})}$, respectively, are given by {[}\ref{I.-Al-Nahhal,-O. TVT}{]}

$\vphantom{}$

$\text{Add}^{(\text{MPA})}=Rd_{f}\left(N_{t}M\right)^{d_{f}}\left(2N_{r}(2d_{f}+1)-1\right)$

\begin{equation}
+KRd_{f}\left(\left(N_{t}M\right)^{d_{f}}-1\right),\label{eq: Add MPA}
\end{equation}

\noindent and

$\vphantom{}$

$\text{Mul}^{(\text{MPA})}=Rd_{f}\left(N_{t}M\right)^{d_{f}}\left(2N_{r}(2d_{f}+1)+Kd_{f}+1\right)$

\begin{equation}
+N_{t}M\left(d_{v}-1\right)\left(KRd_{f}+U\right).\label{eq: Mul MPA}
\end{equation}

\subsection{The SUD Algorithm}

In the SUD algorithm, the cost of (\ref{eq: energy}) is $R(2N_{r}d_{f}-1)$
real additions and $2RN_{r}d_{f}$ real multiplications. The cost
of one possible combination of $j$ and $l$ in (\ref{eq: SUD}) for
$N_{r}$ receive antennas is $N_{r}(4d_{f}+2)-1$ real additions and
$N_{r}(4d_{f}+2)$ real multiplications. The number of possible combinations
between $j$ and $l$ in (\ref{eq: SUD}) varies from one ORE to another
based on the system indicator matrix. Thus, the required number of
real additions and multiplications, $\text{Add}^{(\text{SUD})}$ and
$\text{Mul}^{(\text{SUD})}$, respectively, of the SUD algorithm can
be written as

$\vphantom{}$

$\text{Add}^{(\text{SUD})}=R\left(2N_{r}d_{f}-1\right)$

\begin{equation}
\hspace{2.1cm}+\left(N_{r}\left(4d_{f}+2\right)-1\right)\hspace{-3mm}\sum_{\begin{array}{c}
r=1\\
\grave{U}^{r}\neq0
\end{array}}^{R}\hspace{-3mm}\left(MN_{t}\right)^{\grave{U}^{r}},\label{eq: Add SUD}
\end{equation}

\noindent and

\begin{equation}
\text{Mul}^{(\text{SUD})}=2RN_{r}d_{f}+N_{r}\left(4d_{f}+2\right)\hspace{-3mm}\sum_{\begin{array}{c}
r=1\\
\grave{U}^{r}\neq0
\end{array}}^{R}\hspace{-3mm}\left(MN_{t}\right)^{\grave{U}^{r}}.\label{eq: Mul SUD}
\end{equation}

The summation term in (\ref{eq: Add SUD}) and (\ref{eq: Mul SUD})
depends on the indicator matrix of the system.\footnote{In this paper, the system in (\ref{A}) is considered. Consequently,
the result of the summation term in (\ref{eq: Add SUD}) and (\ref{eq: Mul SUD})
becomes $(MN_{t})^{3}+(MN_{t})^{2}+(MN_{t})^{1}$.}

\subsection{The MSUD Algorithm}

The MSUD algorithm iteratively updates the estimated users messages
of the SUD algorithm at an extra cost of $KUMN_{t}(N_{r}(4d_{f}+2)-1)$
and $KUMN_{t}N_{r}(4d_{f}+2)$ real additions and multiplications,
respectively. Thus, the required number of real additions and multiplications,
$\text{Add}^{(\text{MSUD})}$ and $\text{Mul}^{(\text{MSUD})}$, respectively,
of the MSUD algorithm are given by$^{2}$

$\vphantom{}$

$\text{Add}^{(\text{MSUD})}=R\left(2N_{r}d_{f}-1\right)+\left(N_{r}\left(4d_{f}+2\right)-1\right)$

\begin{equation}
\times\left(\hspace{-1mm}KUMN_{t}+\hspace{-5mm}\sum_{\begin{array}{c}
r=1\\
\grave{U}^{r}\neq0
\end{array}}^{R}\hspace{-5mm}\left(MN_{t}\right)^{\grave{U}^{r}}\hspace{-1mm}\right),\label{eq: Add MSUD}
\end{equation}

\noindent and

$\vphantom{}$

$\text{Mul}^{(\text{MSUD})}=2RN_{r}d_{f}+N_{r}\left(4d_{f}+2\right)$

\begin{equation}
\times\left(\hspace{-1mm}KUMN_{t}+\hspace{-5mm}\sum_{\begin{array}{c}
r=1\\
\grave{U}^{r}\neq0
\end{array}}^{R}\hspace{-5mm}\left(MN_{t}\right)^{\grave{U}^{r}}\hspace{-1mm}\right).\label{eq: Mul MSUD}
\end{equation}

\subsection{The FCSD Algorithm}

The FCSD algorithm visits $\left(MN_{t}\right)^{d_{f}}$ nodes at
the first tree-search level, where each node costs $(N_{r}(4d_{f}+2)-R-2)$
and $N_{r}(4d_{f}+2)$ real additions and multiplications, respectively.
Then, for the rest of $R-1$ levels, the FCSD algorithm visits a fixed
number of nodes at each level according to $\rho_{r}$. Thus, the
required number of real additions and multiplications, $\text{Add}^{(\text{FCSD})}$
and $\text{Mul}^{(\text{FCSD})}$, respectively, of the FCSD algorithm
are given by

$\vphantom{}$

$\text{Add}^{(\text{FCSD})}=R\left(2N_{r}d_{f}-1\right)+\left(N_{r}\left(4d_{f}+2\right)-R-2\right)$

\begin{equation}
\times\left(\left(MN_{t}\right)^{d_{f}}+\sum_{r=2}^{R}\rho_{r-1}\left(MN_{t}\right)^{\grave{U}^{r}}\right),\label{eq: Add FCSD}
\end{equation}

\noindent and

$\vphantom{}$

$\text{Mul}^{(\text{FCSD})}=2RN_{r}d_{f}+N_{r}\left(4d_{f}+2\right)$

\begin{equation}
\times\left(\left(MN_{t}\right)^{d_{f}}+\sum_{r=2}^{R}\rho_{r-1}\left(MN_{t}\right)^{\grave{U}^{r}}\right).\label{eq: Mul FCSD}
\end{equation}

\section{\label{sec:Simulation-Results}Simulation Results and Discussions}

In this section, the proposed decoding algorithms and conventional
MPA decoder in {[}\ref{Z.-Pan,-J.}{]} are assessed using Monte-Carlo
simulations for the SM-SCMA system. The Rayleigh fading channel coefficients
between the transmit and receive antennas for all users are considered
to be perfectly known at the receiver side. An SM-SCMA system of six
users that share four OREs based on (\ref{eq: indicator matrix})
or (\ref{A}) is considered for the assessment (i.e., $U=6$, $R=4$,
$d_{f}=3$ and $d_{v}=2$). Two user spectral efficiencies based on
(\ref{eq:SE}) are considered in the results: $\eta_{u}=3$ bpcu ($N_{t}=4$
and $M=2$) and $\eta_{u}=4$ bpcu ($N_{t}=4$ and $M=4$), and the
$M$-QAM scheme is used in the simulations.

Three MIMO scenarios are studied for each user spectral efficiency:
under-determined MIMO system (e.g., $N_{r}=2$), determined MIMO system
(e.g., $N_{r}=4$) and over-determined MIMO system (e.g., $N_{r}=6$
and $N_{r}=10$). Thus, there are six scenarios within the scope of
this paper (i.e., three MIMO scenarios for each of the two user spectral
efficiencies). It is worth noting that the BER performance of the
conventional MPA decoder for the SM-SCMA converges after five iterations
(i.e., $K=5$) for the considered six scenarios. The following simulation
results are obtained by running at least $10^{5}$ independent realizations.

\begin{figure}
\begin{centering}
\includegraphics[scale=0.39]{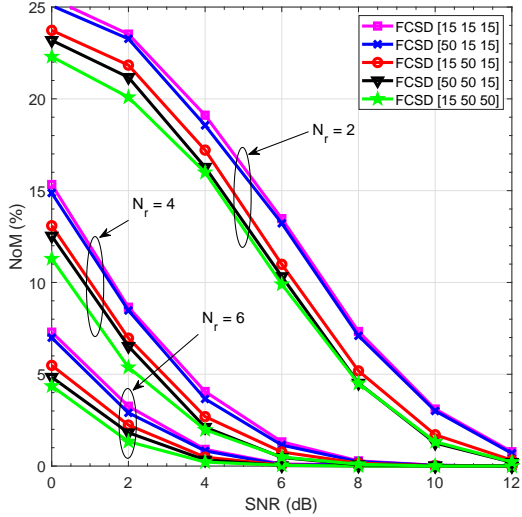}
\par\end{centering}
\caption{\label{fig: NoM eta_3}NoM of different values of $\rho_{r}$ for
$\eta_{u}=3$ bpcu.}
\end{figure}

\subsection{\label{subsec:Parameters-Sensitivity}The Effect of the FCSD Pruned
Radius on BER}

In this subsection, the effect of choosing $\gamma_{r}$ (or $\rho_{r}$)
across the tree-search level on the BER performance for the proposed
FCSD algorithm is studied. In other words, we need to know which level
of the tree-search has a great effect on the BER performance to increase/decrease
the number of visited nodes. As seen from (\ref{eq: CDF}), as $\gamma_{r}$
increases, the probability of not missing the optimal solution (i.e.,
MPA solution) increases. The question that arises is which level has
a significant effect on the probability in (\ref{eq: CDF}). To answer
this question, let us define the number of misses (NoM) as the number
of times that the FCSD algorithm misses the MPA solution. The NoM
can be used as an indicator to study the effect of selecting $\gamma_{r}$
at each level, taking into account that a small value of the NoM reflects
an acceptable BER performance and vice versa. Thus, the NoM can be
formulated as

\begin{equation}
\text{NoM}=\mathbb{E}\left\{ \sum_{u=1}^{U}\mathcal{P}\left(\left\{ \hat{\mathbf{c}}_{m}^{u},\hat{n_{t}}^{u}\right\} |_{\text{FCSD}}\neq\left\{ \hat{\mathbf{c}}_{m}^{u},\hat{n_{t}}^{u}\right\} |_{\text{MPA}}\right)\right\} ,\label{eq: NoM}
\end{equation}

\noindent where $\mathcal{P}(\centerdot)$ in (\ref{eq: NoM}) equals
$1$ or $0$ when the estimated messages of the $u$-th user using
the FCSD and MPA decoders are different or the same, respectively.

To study the effect of $\gamma_{r}$, the FCSD algorithm with $[\rho_{1}\,\,\rho_{2}\,\,\rho_{3}]=$
$[15\,\,15\,\,15]$ is assumed to be the baseline of this study for
the three MIMO scenarios of $\eta_{u}=3$ bpcu. To note the effect
of $\rho_{r}$ on the BER performance of each level, we increase the
number of survived nodes of only one level at a time, while the number
of survived nodes of the rest of levels is kept the same. As such,
to see the effect of the first level (i.e., $\rho_{1}$) on the BER
performance compared to the baseline, we notice the improvement in
the NoM when $\rho_{1}=50$ and $\rho_{2}=\rho_{3}=15$ (i.e., $[\rho_{1}\,\,\rho_{2}\,\,\rho_{3}]=$
$[50\,\,15\,\,15]$). It should be noted that these numbers are arbitrarily
chosen to study the effect of $\rho_{r}$ on the BER performance.
Next, we do the same thing for the second level (i.e., $[\rho_{1}\,\,\rho_{2}\,\,\rho_{3}]=$
$[15\,\,50\,\,15]$) and notice the improvement in the NoM. From Fig.
\ref{fig: NoM eta_3}, the improvement in NoM for the first level
is negligible compared to the improvement in the NoM obtained from
increasing the survived nodes in the second level. Consequently, the
second level has a greater effect on BER performance compared to the
first level.

\begin{figure}
\begin{centering}
\includegraphics[scale=0.39]{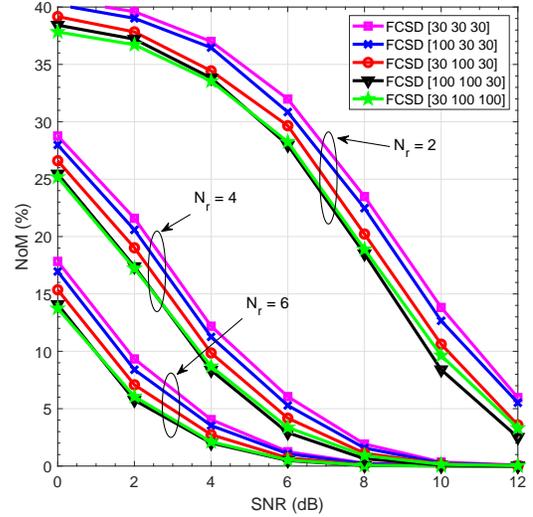}
\par\end{centering}
\caption{\label{fig: NoM eta_4}NoM of different values of $\rho_{r}$ for
$\eta_{u}=4$ bpcu.}
\end{figure}

It should be noted that $\rho_{3}$ can not be greater than $\rho_{2}$
since both belong to SES, as in Fig. \ref{fig:Tree-search-of-the}.
Therefore, to continue the study for $\rho_{1}$ and $\rho_{3}$,
let us consider $[\rho_{1}\,\,\rho_{2}\,\,\rho_{3}]=$ $[15\,\,50\,\,15]$
as a new baseline for  comparison. Compared to the new baseline: first,
the number of survived nodes of the first level is increased to be
50 (i.e., $[\rho_{1}\,\,\rho_{2}\,\,\rho_{3}]=$ $[50\,\,50\,\,15]$),
then we do the same thing for the third level (i.e., $[\rho_{1}\,\,\rho_{2}\,\,\rho_{3}]=$
$[15\,\,50\,\,50]$) and notice the improvement in the NoM. As depicted
in Fig. \ref{fig: NoM eta_3}, the improvement in NoM from the third
level is larger than the improvement obtained from the first level,
compared to the new baseline.

By taking an in-depth look at Fig. \ref{fig: NoM eta_3}, one can
observe that the increase in the number of survived nodes at the second
level provides better NoM improvements, compared to the increase in
the number of survived nodes at any other level. The reason is that
only part of users share the upper levels of FES; thus, the distance
metric nodes at FES levels do not represent all users. On the other
hand, the nodes at SES levels include the distance metrics of all
users, which significantly affects the BER performance. Hence, the
second level has the highest effect on the BER performance, then the
third level, and finally the first level. In essence, increasing the
number of survived nodes at the SES levels is more effective than
at the FES levels, especially the upper levels of the SES. It should
be noted that the conclusion drawn from this study is independent
on the structure of $F$ in (\ref{eq: indicator matrix}).

Fig. \ref{fig: NoM eta_4} shows the effect of $\rho_{r}$ on the
BER performance in terms of NoM for $\eta_{u}=4$ bpcu. In these scenarios,
the FCSD algorithm with $[\rho_{1}\,\,\rho_{2}\,\,\rho_{3}]=$ $[30\,\,30\,\,30]$
is considered. As discussed for $\eta_{u}=3$ bpcu, $\rho_{2}$ provides
significant improvements in the NoM. On the other hand, $\rho_{1}$
and $\rho_{3}$ provide almost the same improvements for the three
scenarios depicted in Fig. \ref{fig: NoM eta_4}. In other words,
there is no preference for increasing the number of survived nodes
at these two levels from the NoM perspective. However, it is preferable
to increase $\rho_{3}$ rather than $\rho_{1}$ from the decoding
complexity point of view, as seen from (\ref{eq: Add FCSD}) and (\ref{eq: Mul FCSD}).
This means that increasing $\rho_{3}$ results in a lower increase
in the decoding complexity compared with the increase of $\rho_{1}$.

Finally, increasing the number of survived nodes at the lower tree-search
levels has a better effect on the BER performance or/and decoding
complexity. It is worth noting that the number of survived nodes at
the first levels should be empirically chosen to avoid the early dropping
of the MPA solution. Empirically, the FCSD algorithm with $[35\,\,70\,\,50]$
and $[110\,\,320\,\,300]$ provides near MPA BER performances (i.e.,
NoM close to zero) for $\eta_{u}=3$ bpcu and $\eta_{u}=4$ bpcu,
respectively.

\subsection{BER Performance Assessment}

In this subsection, the BER performance of the proposed decoders is
compared with the conventional MPA versus different values of SNR
for all six scenarios. The proposed MSUD algorithm and conventional
MPA converges at four and five iterations, respectively (i.e., $K=4$
for MSUD and $K=5$ for MPA). Moreover, $K=1$ is provided for the
MSUD and MPA to highlight the improvement in the BER performance when
using the value of $K$ at the convergence for both algorithms.

\begin{figure*}
\hspace*{-3mm}\subfloat[\label{fig: N_r_2 eta 3}$N_{r}=2$.]{\begin{centering}
\includegraphics[scale=0.32]{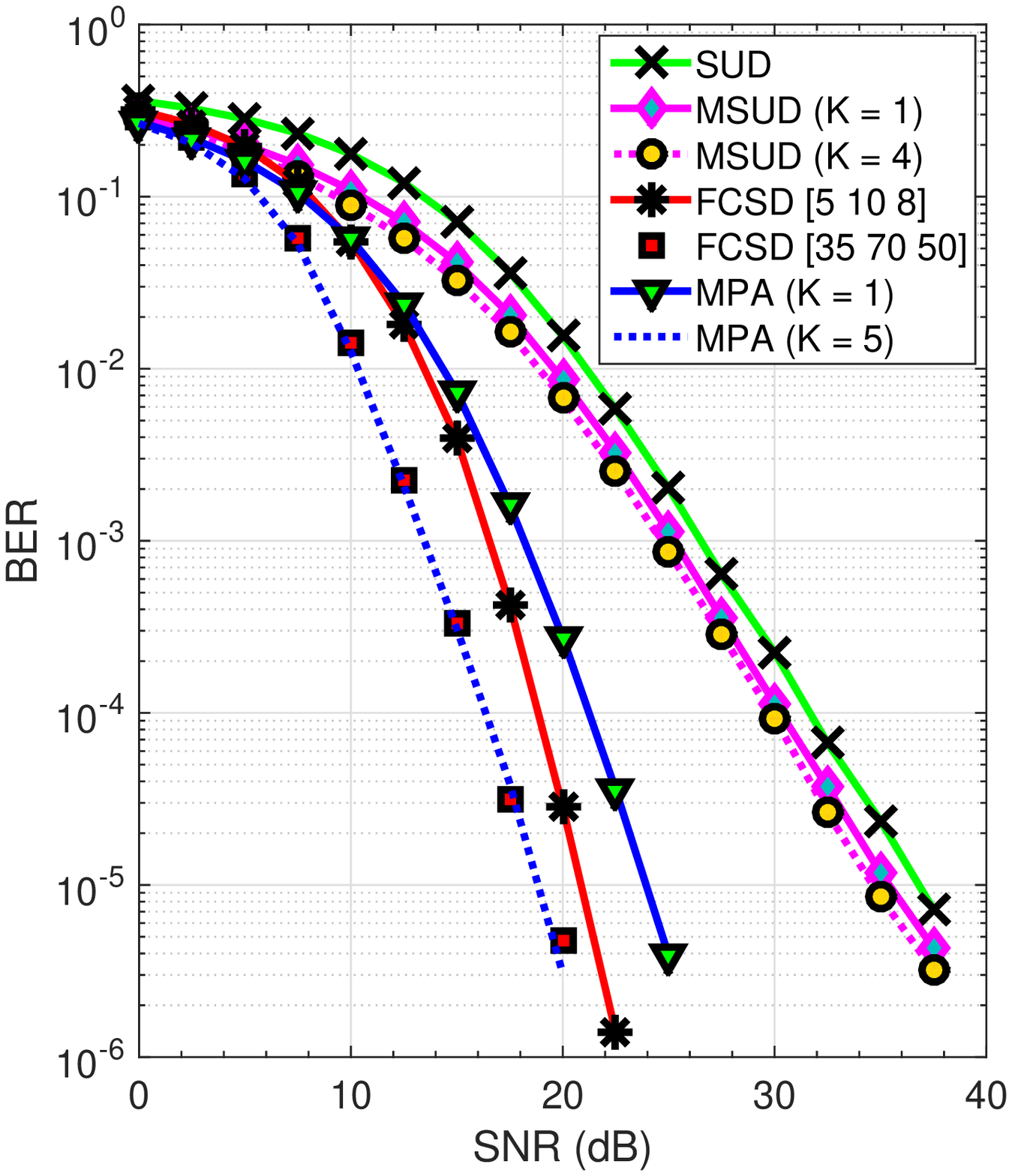}
\par\end{centering}
}\hspace*{-8mm}\subfloat[\label{fig: N_r_4 eta 3}$N_{r}=4$.]{\begin{centering}
\includegraphics[scale=0.32]{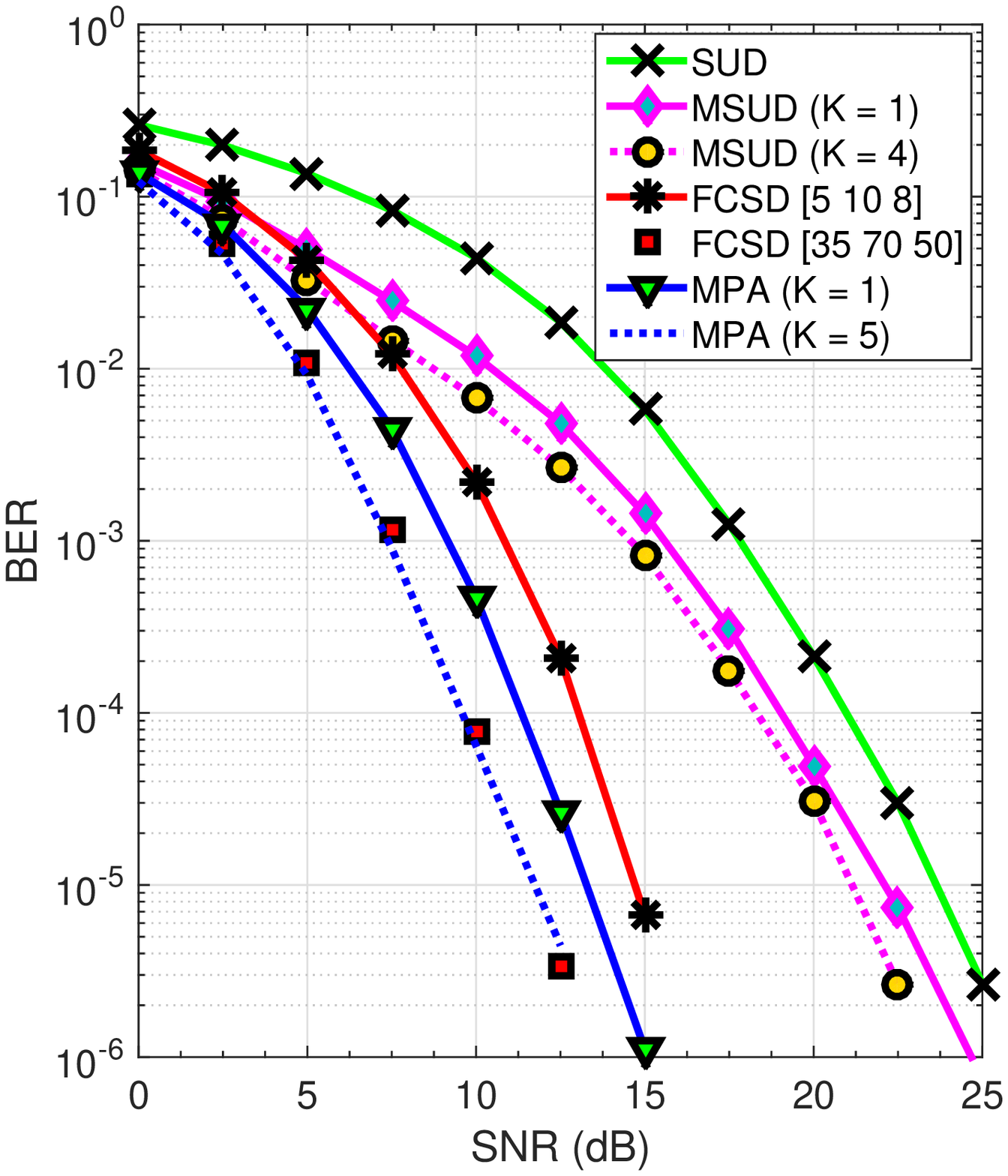}
\par\end{centering}
}\hspace*{-8mm}\subfloat[\label{fig: N_r_6 eta 3}$N_{r}=6$.]{\begin{centering}
\includegraphics[scale=0.32]{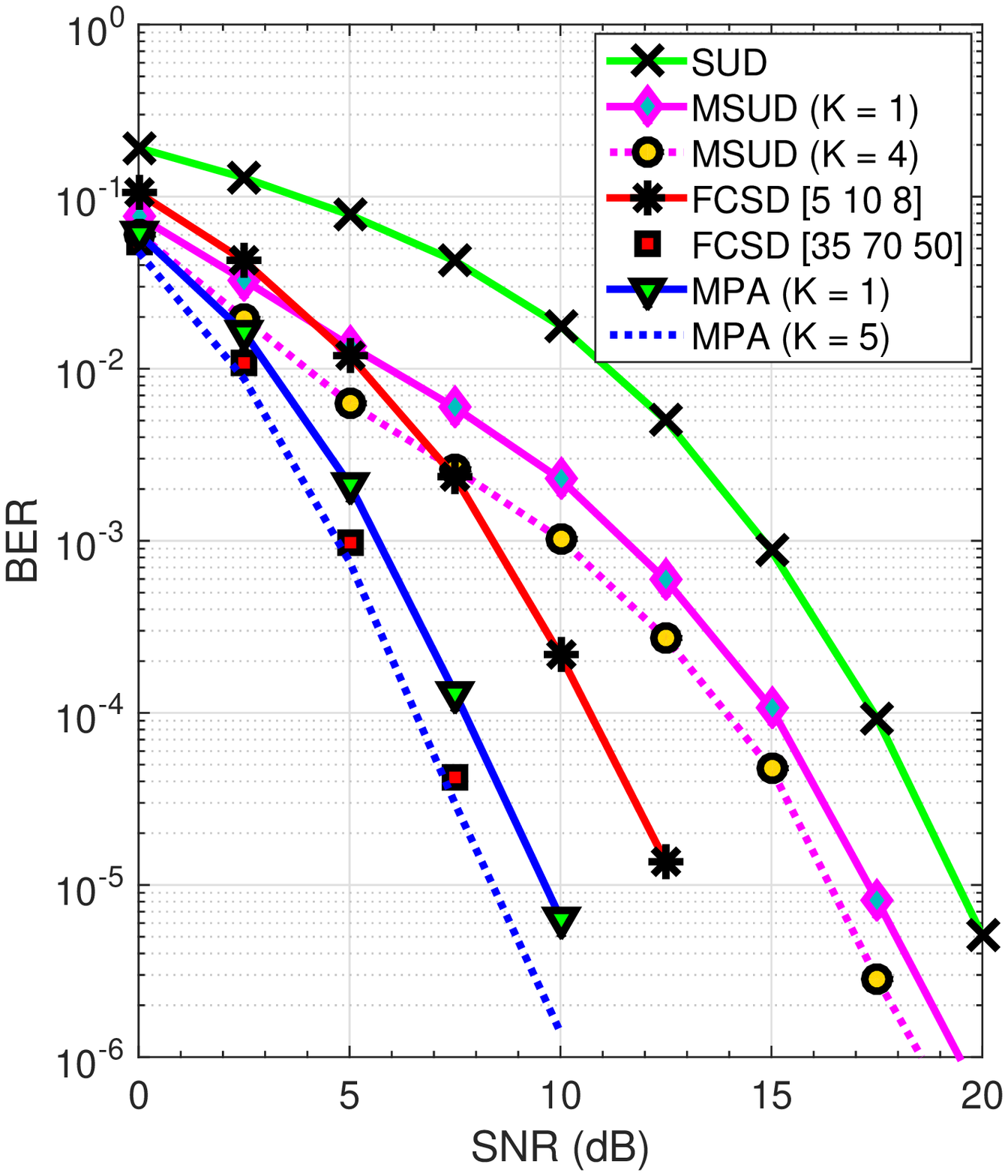}
\par\end{centering}
}\hspace*{-8mm}\subfloat[\label{fig: N_r_10 eta 3}$N_{r}=10$.]{\begin{centering}
\includegraphics[scale=0.32]{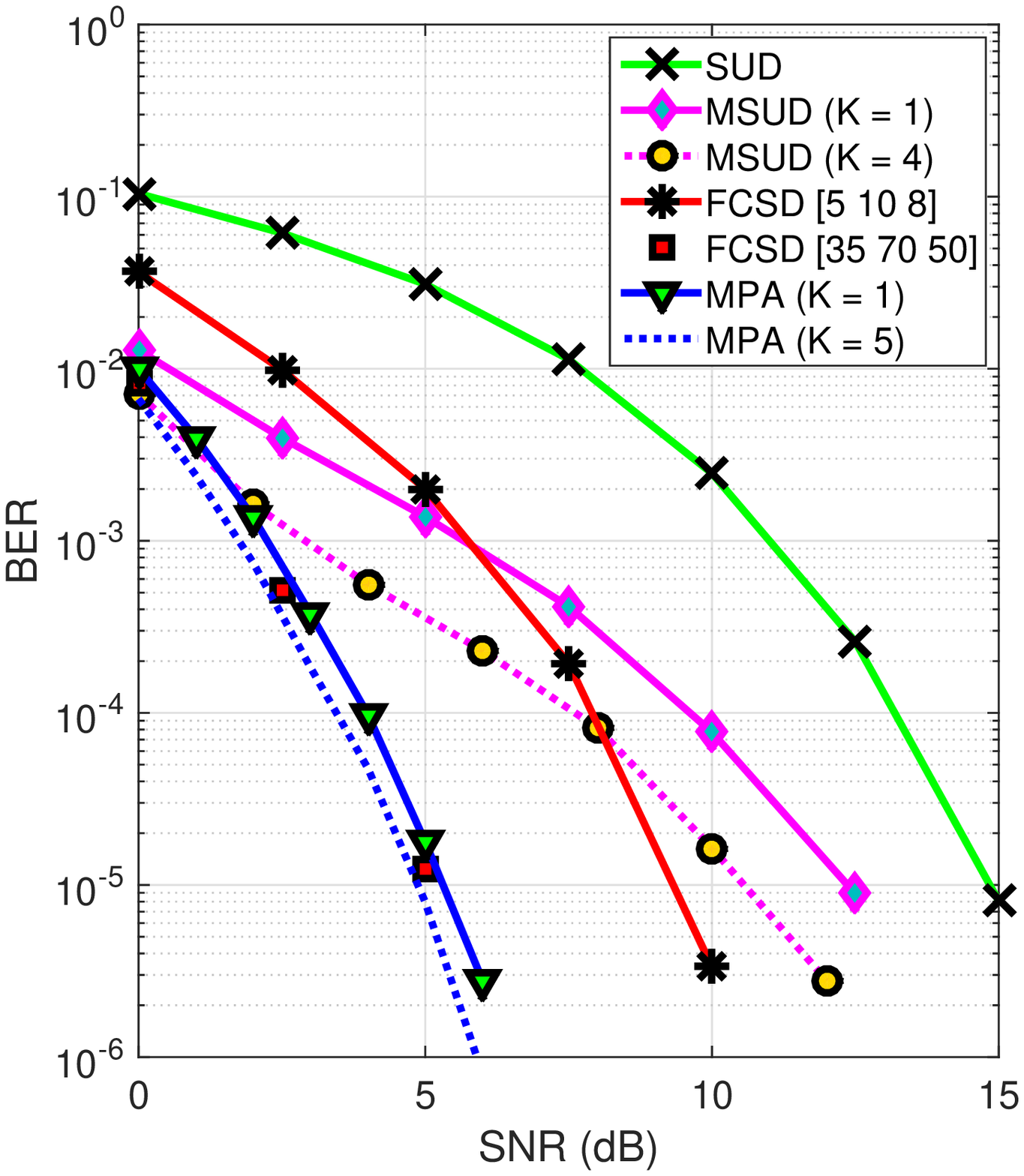}
\par\end{centering}
}

\caption{\label{fig:BER-eta_3}BER performance of different SM-SCMA decoders
for $N_{r}\times4$ MIMO with $M=2$ for each user (i.e., $\eta_{u}=3$
bpcu).}
\end{figure*}

Fig. \ref{fig:BER-eta_3} depicts the BER performance of the proposed
and MPA decoders for $\eta_{u}=3$ bpcu in different three MIMO scenarios
(i.e., $N_{r}=2$, $4$, $6$ and $10$). As mentioned in Subsection
\ref{subsec:Parameters-Sensitivity} and as seen from this figure,
the proposed FCSD algorithm with $[\rho_{1}\,\,\rho_{2}\,\,\rho_{3}]=$
$[35\,\,70\,\,50]$ provides a very similar BER performance as MPA.
The FCSD algorithm with $[\rho_{1}\,\,\rho_{2}\,\,\rho_{3}]=$ $[5\,\,10\,\,8]$
is depicted in Fig. \ref{fig:BER-eta_3} to show that the FCSD can
provide a flexible trade-off between the BER performance and decoding
complexity. It is also shown that the proposed SUD provides an acceptable
BER performance with a considerable degradation in the BER performance
of the MPA. The MSUD with $K=1$ and $K=4$ both provide a considerable
improvement in the SUD BER performance.

\begin{figure*}
\hspace*{-3mm}\subfloat[\label{fig: N_r_2 eta 4}$N_{r}=2$.]{\begin{centering}
\includegraphics[scale=0.32]{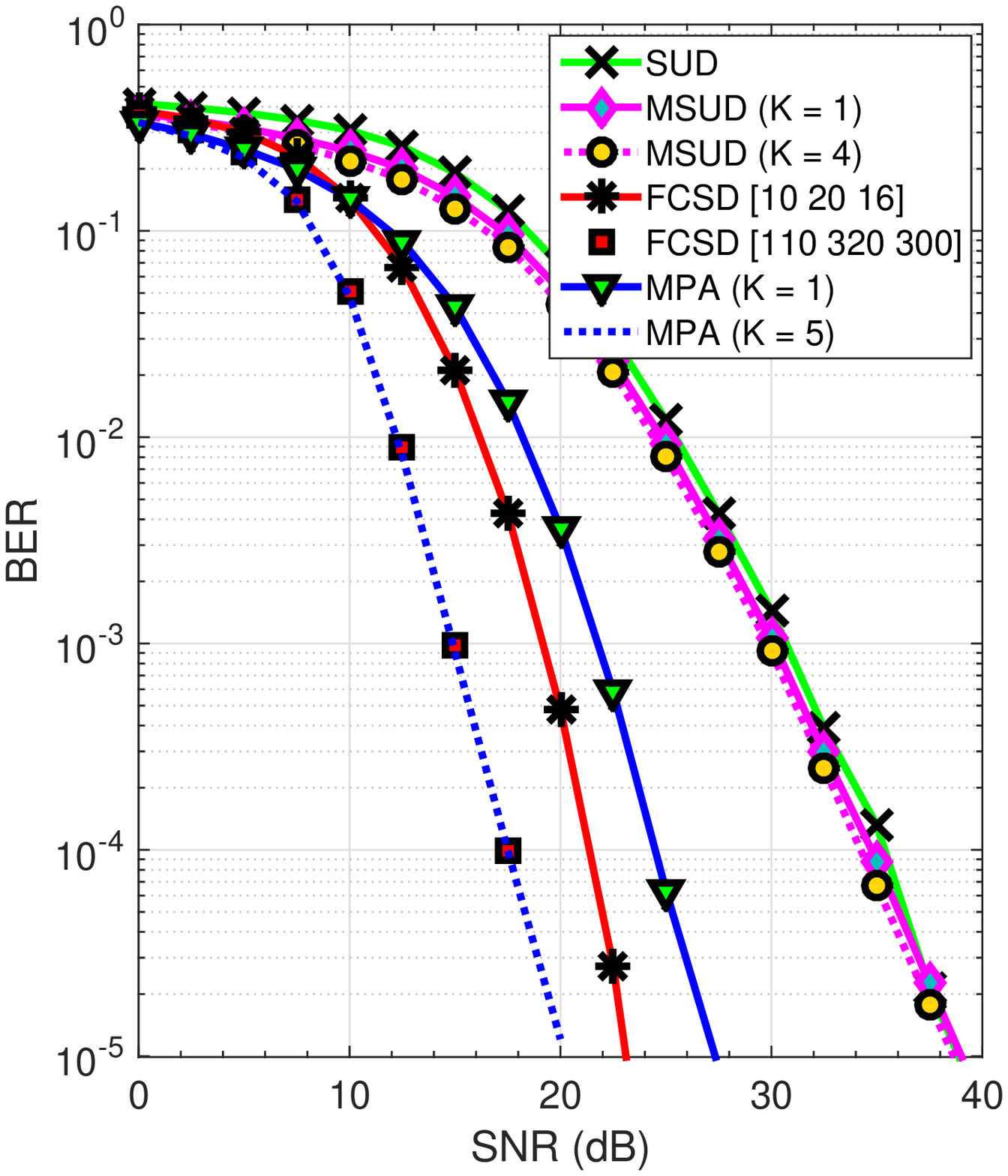}
\par\end{centering}
}\hspace*{-8mm}\subfloat[\label{fig: N_r_4 eta 4}$N_{r}=4$.]{\begin{centering}
\includegraphics[scale=0.32]{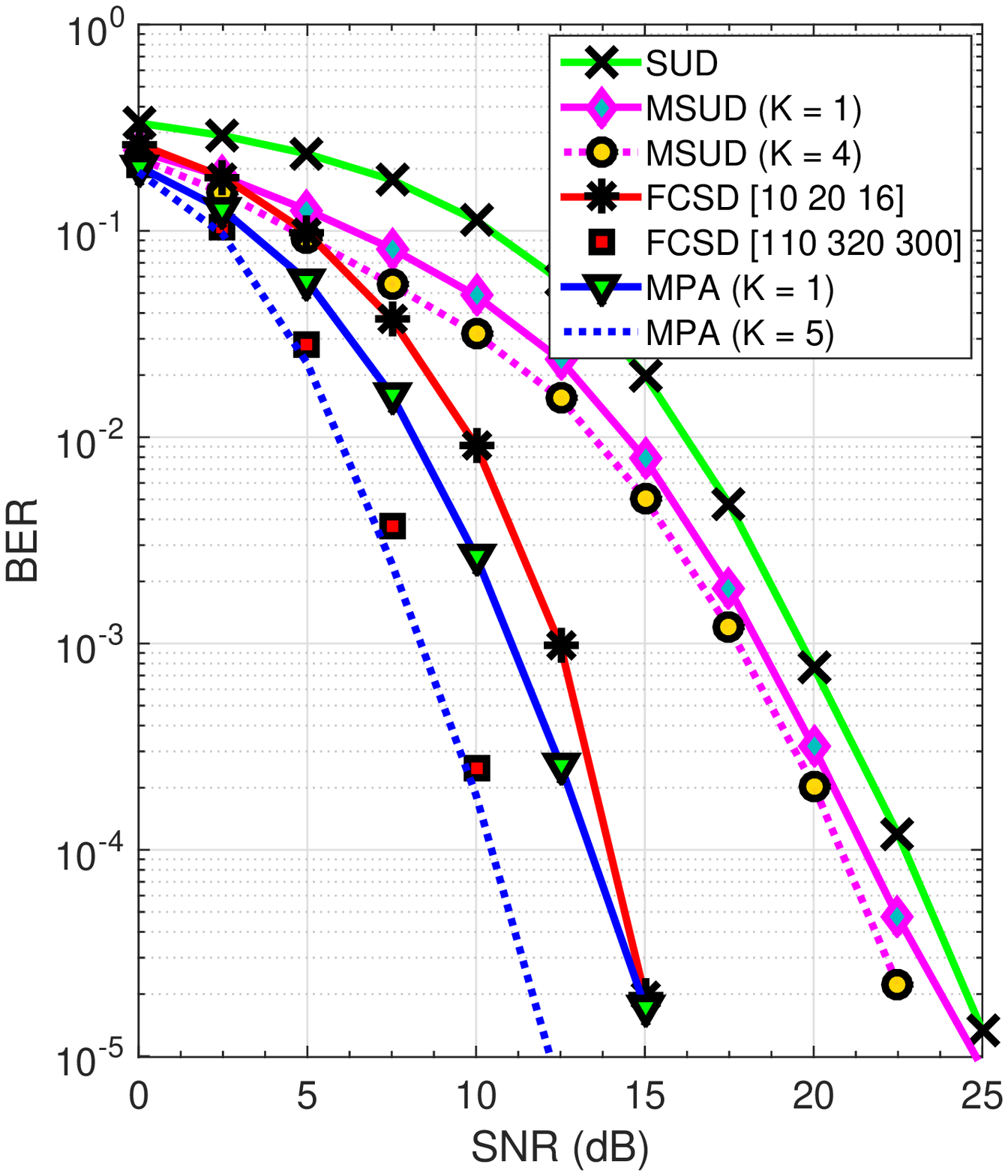}
\par\end{centering}
}\hspace*{-8mm}\subfloat[\label{fig: N_r_6 eta 4}$N_{r}=6$.]{\begin{centering}
\includegraphics[scale=0.32]{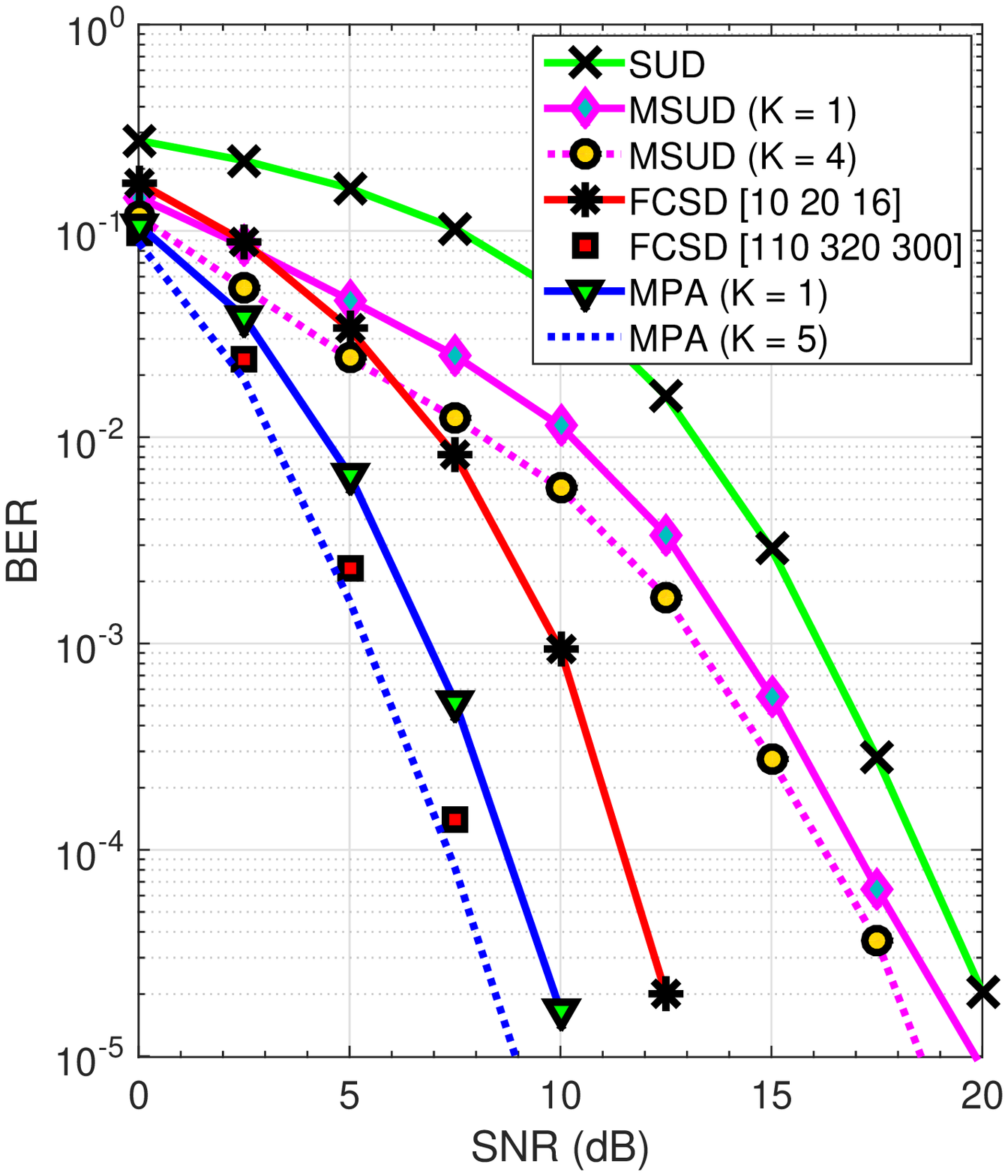}
\par\end{centering}
}\hspace*{-8mm}\subfloat[\label{fig: N_r_10 eta 4}$N_{r}=10$.]{\begin{centering}
\includegraphics[scale=0.32]{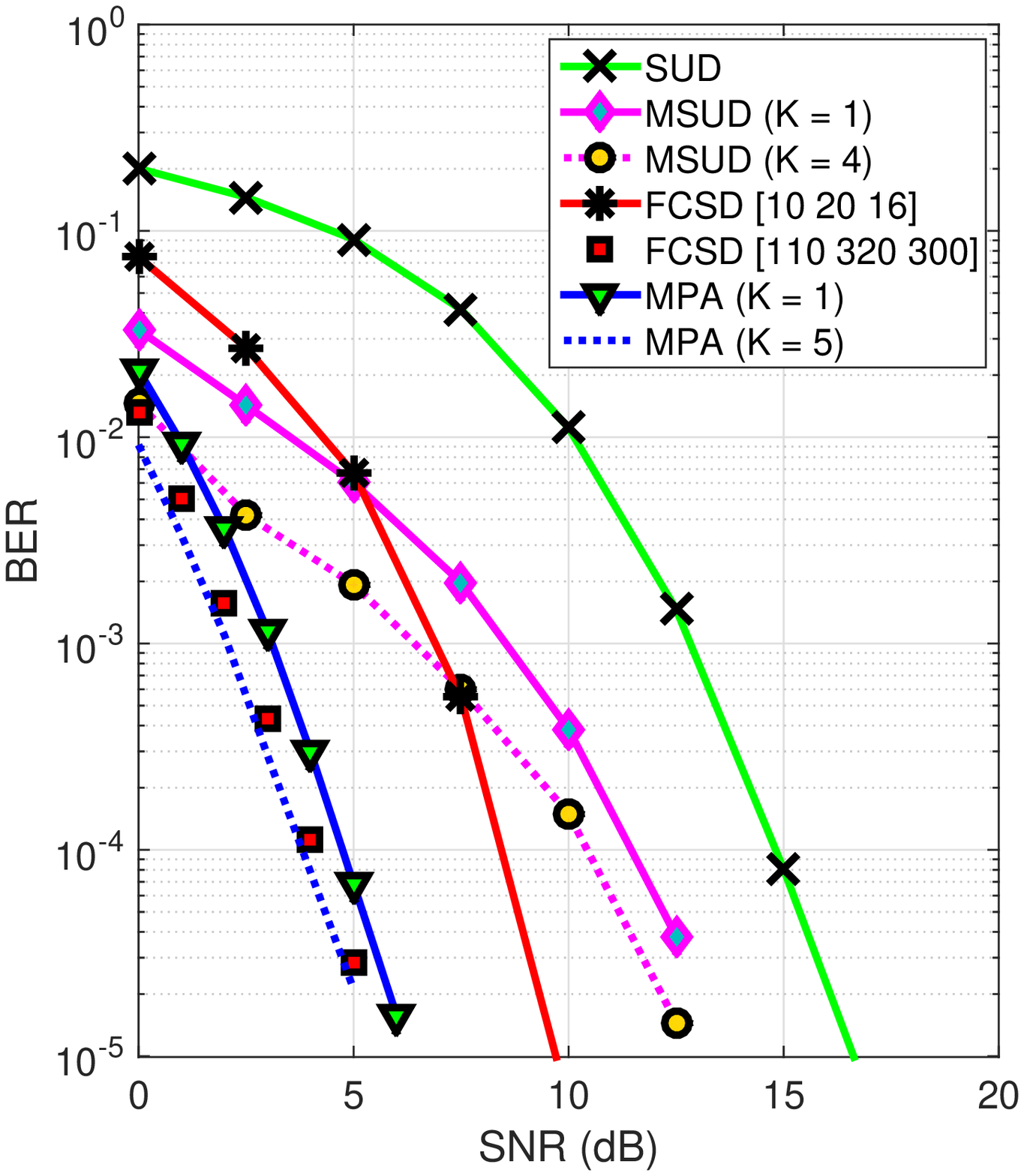}
\par\end{centering}
}

\caption{\label{fig:BER-eta_4}BER performance of different SM-SCMA decoders
for $N_{r}\times4$ MIMO with $M=4$ for each user (i.e., $\eta_{u}=4$
bpcu).}
\end{figure*}

Fig. \ref{fig:BER-eta_4} shows the BER performance of the proposed
and MPA decoders for $\eta_{u}=4$ bpcu in three different MIMO scenarios
(i.e., $N_{r}=2$, $4$ and $6$). Here, the value of $[\rho_{1}\,\,\rho_{2}\,\,\rho_{3}]$
of the proposed FCSD algorithm is modified to be $[110\,\,320\,\,300]$
to provide a very similar BER performance as MPA. Same as the findings
of Fig. \ref{fig:BER-eta_3}, the SUD algorithm yields an acceptable
BER performance, while the MSUD algorithm significantly improves the
BER performance of the SUD algorithm, as seen in Fig. \ref{fig:BER-eta_4}.

\subsection{Decoding Complexity Assessment}

In this subsection, the decoding complexity of the proposed and MPA
decoders are compared in terms of the required number of real additions
and multiplications, based on the deduced equations mentioned in Section
\ref{sec:Complexity-Analysis}.

\begin{figure}
\begin{centering}
\includegraphics[scale=0.41]{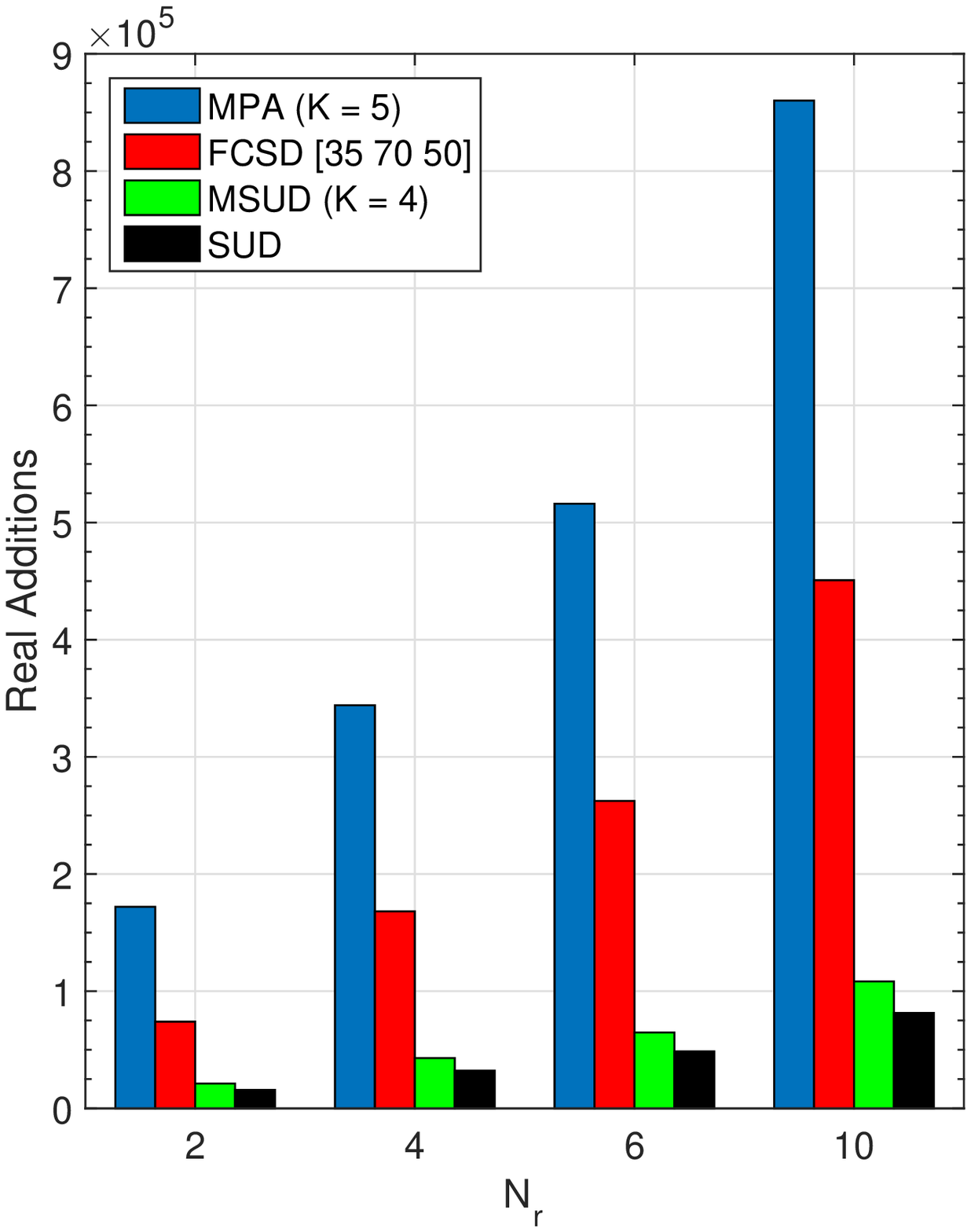}
\par\end{centering}
\caption{\label{fig:Real-additions-eta 3}Real additions comparison of different
SM-SCMA decoders for $\eta_{u}=3$ bpcu.}
\end{figure}

\begin{figure}
\begin{centering}
\includegraphics[scale=0.41]{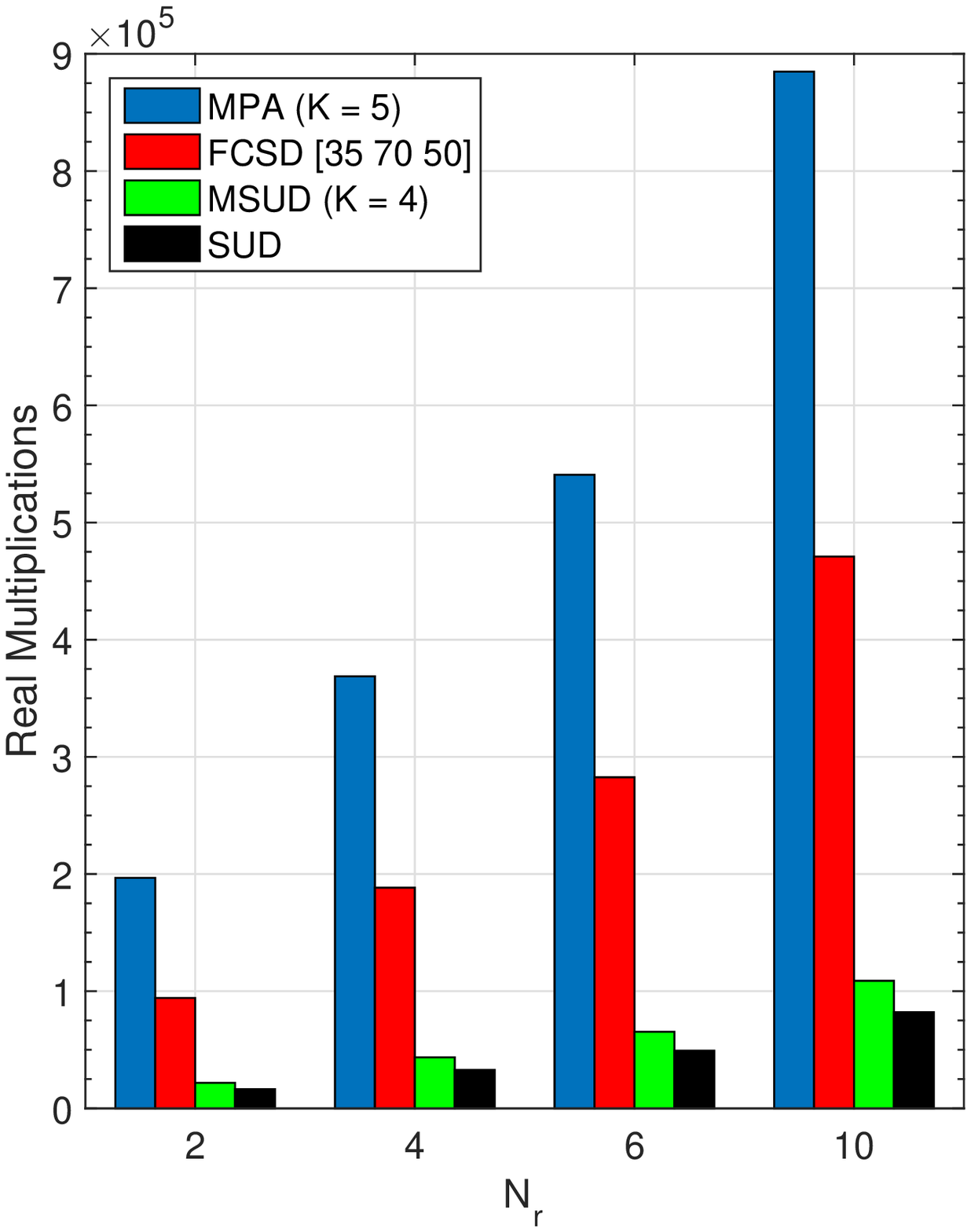}
\par\end{centering}
\caption{\label{fig:Real-multiplications-eta 3}Real multiplications comparison
of different SM-SCMA decoders for $\eta_{u}=3$ bpcu.}
\end{figure}

\begin{figure}
\begin{centering}
\includegraphics[scale=0.41]{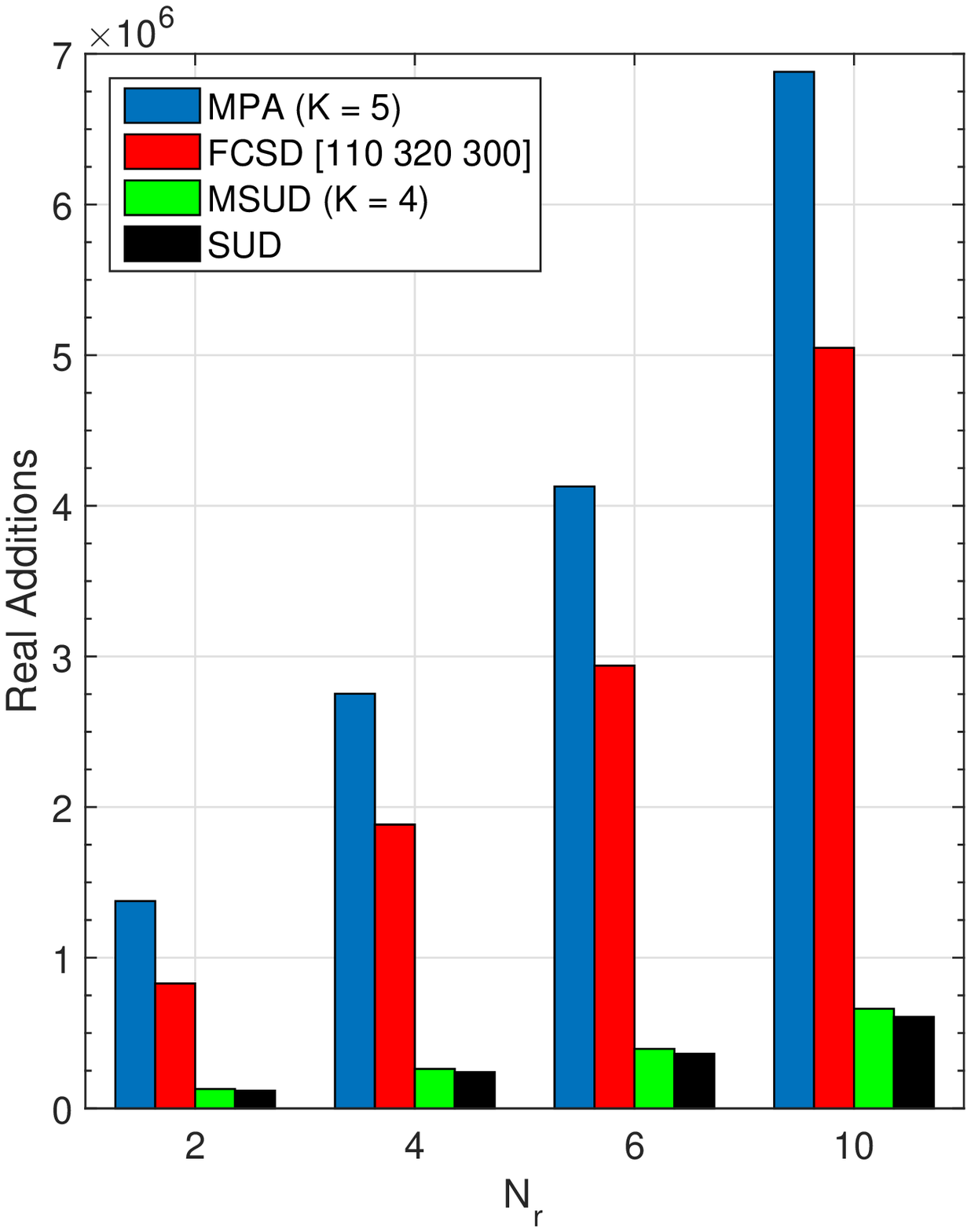}
\par\end{centering}
\caption{\label{fig:Real-additions-eta 4}Real additions comparison of different
SM-SCMA decoders for $\eta_{u}=4$ bpcu.}
\end{figure}

\begin{figure}
\begin{centering}
\includegraphics[scale=0.41]{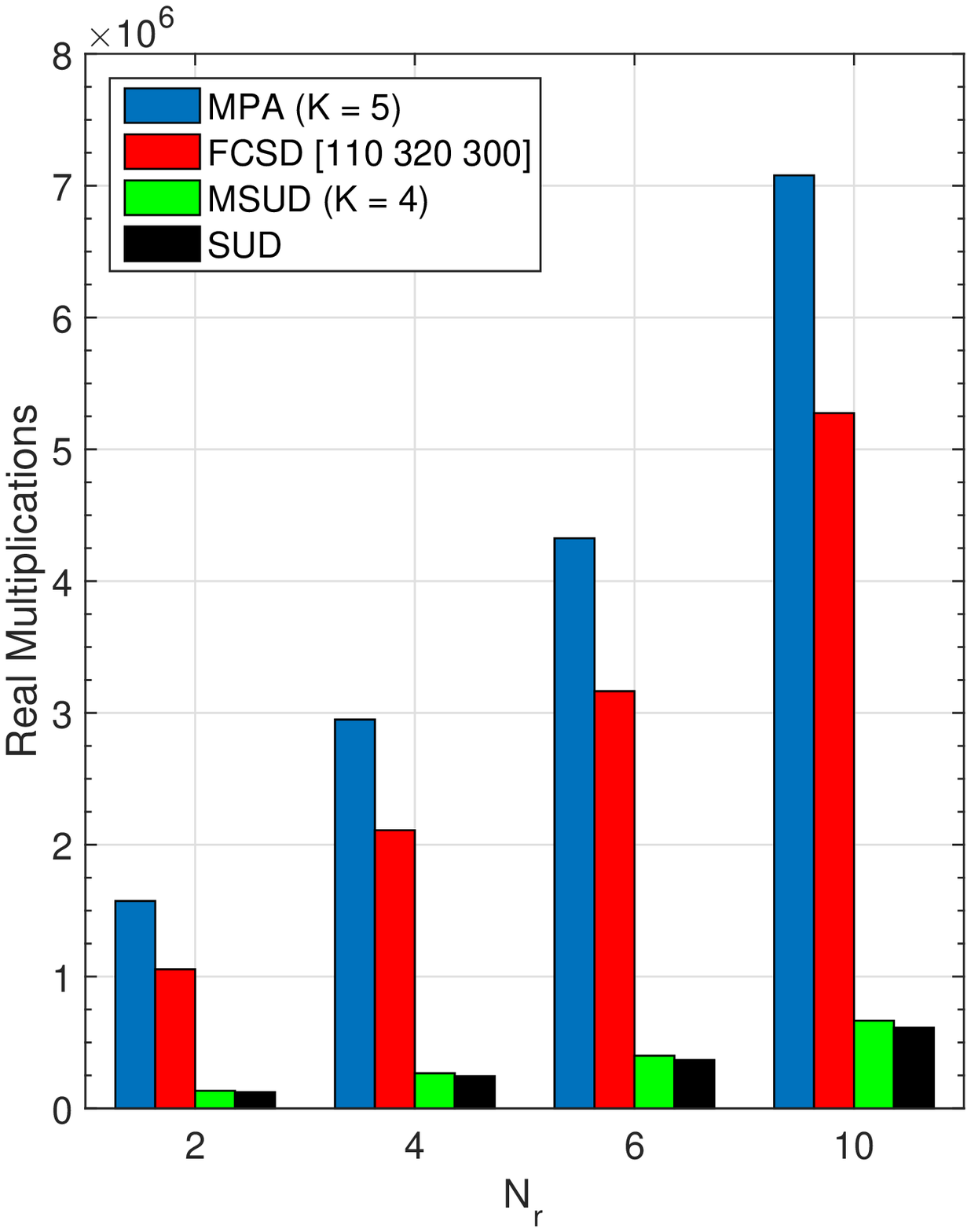}
\par\end{centering}
\caption{\label{fig:Real-multiplications-eta 4}Real multiplications comparison
of different SM-SCMA decoders for $\eta_{u}=4$ bpcu.}
\end{figure}

Figs. \ref{fig:Real-additions-eta 3} and \ref{fig:Real-multiplications-eta 3}
show the required number of real additions and multiplications, respectively,
for $\eta_{u}=3$ bpcu for the three MIMO scenarios. On the other
hand, Figs. \ref{fig:Real-additions-eta 4} and \ref{fig:Real-multiplications-eta 4}
depict the required number of real additions and multiplications,
respectively, for $\eta_{u}=4$ bpcu for the three MIMO scenarios
(i.e., $N_{r}=2$, $4$, $6$ and $10$). It can be inferred from
all these figures that the proposed SUD algorithm provides the lowest
decoding complexity and is significantly low when compared with the
MPA and FCSD algorithms. The proposed MSUD algorithm slightly increases
the decoding complexity compared with the SUD algorithm; however,
its decoding complexity is still very low when compared with the MPA.
Finally, although the complexity of the FCSD algorithm is higher when
compared with the SUD and MSUD algorithms, it is still significantly
lower when compared with MPA.

\subsection{Discussions}

The proposed SUD and MSUD algorithms provide more than a $90\%$ reduction
in the decoding complexity compared to MPA at the expense of BER performance
loss. However, this BER performance loss decreases as $N_{r}$ increases.
For example, the BER loss is around 10 dB and 5.5 dB for $N_{r}=6$
and $10$, respectively, as seen from Figs. \ref{fig:BER-eta_3}(c),
\ref{fig:BER-eta_3}(d), \ref{fig:BER-eta_4}(c) and \ref{fig:BER-eta_4}(d).
Moreover, for good quality links (i.e., moderate and high SNR), these
algorithms provide a good and acceptable BER performance. For instance,
the BER performance is around $10^{-5}$ at less than 20 dB and 15
dB for $N_{r}=6$ and $10$, respectively, which is an acceptable
value.

Thus, the proposed SUD and MSUD algorithms are important, as they
exhibit a significant reduction in the decoding complexity while providing
an acceptable BER performance in the case of a good quality link and/or
with a higher number of receive antennas (which is feasible in the
uplink scenario). Furthermore, the proposed FCSD algorithm provides
a near-optimum BER performance with reduced decoding complexity. This
variety of proposed decoders may fit a wide range of possible candidate
applications in practice.

Furthermore, the three proposed decoder concepts may be adapted to
decode the RGSM-SCMA {[}\ref{I.-Al-Nahhal,-O. TVT}{]} signals. However,
the formulation and settings of the algorithms need further investigation.
In other words, since the RGSM-SCMA system activates more than one
antenna associated with rotational angles, a suitable mathematical
reformulation for the proposed algorithms is needed.

\section{\label{sec:Conclusions}Conclusions}

This paper proposes three different low-complexity decoding algorithms,
for the first time, for the uplink SM-SCMA system. The proposed SUD
algorithm is a non-iterative algorithm that provides a benchmark for
the decoding complexity at the expense of the BER performance which
is still acceptable for some possible practical applications under
certain environments and settings. The degradation of its BER performance
comes from using only some of the available OREs in estimating the
users messages. The proposed MSUD algorithm is an iterative algorithm
that considerably improves the BER performance of the SUD algorithm,
with the cost of a slight increase in the decoding complexity. The
MSUD algorithm uses all available OREs to decode the users messages.
The proposed FCSD algorithm provides a close BER performance as MPA
with a considerable reduction in the decoding complexity. Unlike the
MPA, the proposed FCSD algorithm supports parallel hardware implementation.
These proposed algorithms can fit a wide range of possible practical
applications with specific requirements for both operation and hardware
implementation. The mathematical formulation, complexity analysis
for all proposed algorithms, and simulation results are provided to
support these findings. As a potential direction, the proposed algorithm
may be extended to decode the RGSM-SCMA signals.

\begin{IEEEbiography}{}
\end{IEEEbiography}


\begin{thebibliography}{10}
\bibitem{key-2}\label{W.-Shin,-M.}W. Shin, M. Vaezi, B. Lee, D.
J. Love, J. Lee, and H. V. Poor, \textquotedblleft Non-orthogonal
multiple access in multi-cell networks: Theory, performance, and practical
challenges,\textquotedblright{} \textit{IEEE Commun. Mag.}, vol. 55,
no. 10, pp. 176--183, Oct. 2017.

\bibitem{key-8}\label{A. Yadav}A. Yadav and O. A. Dobre, ``All
technologies work together for good: A glance at future mobile networks,''
in \textit{IEEE Wireless Commun}., vol. 25, no. 4, pp. 10-16, Aug.
2018.

\bibitem{key-1}\label{M.-Mohammadkarimi,-M.}M. Mohammadkarimi, M.
A. Raza, and O. A. Dobre, \textquotedblleft Signature-based nonorthogonal
massive multiple access for future wireless networks: Uplink massive
connectivity for machine-type communications,\textquotedblright{}
\emph{IEEE Veh. Technol. Mag.}, vol. 13, no. 4, pp. 40--50, Dec.
2018.

\bibitem{key-2}\label{S.-M.-R. Islam}S. M. R. Islam, N. Avazov,
O. A. Dobre, and K. S. Kwak, \textquotedblleft Power-domain non-orthogonal
multiple access (NOMA) in 5G systems: Potentials and challenges,\textquotedblright{}
\textit{IEEE Commun. Surv. Tuts.}, vol. 19, no. 2, pp. 721-742, Oct.
2016.

\bibitem{key-3}\label{Z.-Ding-et}Z. Ding, X. Lei, G. K. Karagiannidis,
R. Schober, J. Yuan, and V. Bhargava, \textquotedblleft A survey on
non-orthogonal multiple access for 5G networks: Research challenges
and future trends,\textquotedblright{} \textit{IEEE J. Sel. Areas
Commun.}, vol. 35, no. 10, pp. 2181 --2195, Oct. 2017.

\bibitem{key-4}\label{H.-Nikopour-and}H. Nikopour and H. Baligh,
\textquotedblleft Sparse code multiple access,\textquotedblright{}
in \textit{Proc. IEEE Int. Symposium on Personal Indoor and Mobile
Radio Commun. (PIMRC)}, Sep. 2013, pp. 332--336.

\bibitem{key-5}\label{M.-Taherzadeh-et}M. Taherzadeh, H. Nikopour,
A. Bayesteh, and H. Baligh, \textquotedblleft SCMA codebook design,\textquotedblright{}
in \textit{Proc. IEEE Veh. Technol. Conf. (VTC Fall)}, Sep. 2014,
pp. 1--5.

\bibitem{key-3}\label{D.-Cai,-P.}D. Cai, P. Fan, X. Lei, Y. Liu,
and D. Chen, \textquotedblleft Multi-dimensional SCMA codebook design
based on constellation rotation and interleaving,\textquotedblright{}
in \textit{Proc. IEEE Veh. Technol. Conf. (VTC Spring)}, May 2016,
pp. 1--5.

\bibitem{key-1}\label{M.-Vameghestahbanati,-I.}M. Vameghestahbanati,
I. D. Marsland, R. H. Gohary, and H. Yanikomeroglu, \textquotedblleft Multidimensional
constellations for uplink SCMA systems - A comparative study,\textquotedblright{}
\textit{arXiv}, Apr. 2018. {[}Online{]}. Available: http://arxiv.org/abs/1804.05814.

\bibitem{key-6}\label{H.-Mu,-Z.}H. Mu, Z. Ma, M. Alhaji, P. Fan,
and D. Chen, \textquotedblleft A fixed low complexity message pass
algorithm detector for up-link SCMA system,\textquotedblright{} \textit{IEEE
Wireless Commun}.\textit{ Lett}.\textit{,} vol. 4, no. 6, pp. 585--588,
Dec. 2015.

\bibitem{key-1}\label{L.-Yang,-Y.}L. Yang, Y. Liu, and Y. Siu, \textquotedblleft Low
complexity message passing algorithm for SCMA system,\textquotedblright{}
\textit{IEEE Commun. Lett.}, vol. 20, no. 12, pp. 2466--2469, Dec.
2016.

\bibitem{key-1}\label{J.-Dai,-K.}J. Dai, K. Niu, C. Dong, and J.
Lin, \textquotedblleft Improved message passing algorithms for sparse
code multiple access,\textquotedblright{} \textit{IEEE Trans. Veh.
Technol.}, vol. 66, no. 11, pp. 9986--9999, Nov. 2017.

\bibitem{key-2}\label{L.-Yang,-X.}L. Yang, X. Ma, and Y. Siu, \textquoteleft \textquoteleft Low
complexity MPA detector based on sphere decoding for SCMA,\textquoteright \textquoteright{}
\textit{IEEE Commun. Lett.}, vol. 21, no. 8, pp. 1855--1858, Aug.
2017.

\bibitem{key-3}\label{M.-Vameghestahbanati,-E. bdeer}M. Vameghestahbanati,
E. Bedeer, I. Marsland, R. H. Gohary, and H. Yanikomeroglu, \textquoteleft \textquoteleft Enabling
sphere decoding for SCMA,\textquoteright \textquoteright{} \textit{IEEE
Commun. Lett.}, vol. 21, no. 12, pp. 2750--2753, Dec. 2017.

\bibitem{key-4}\label{L.-Li,-J. Modified SD}L. Li, J. Wen, X. Tang,
and C. Tellambura, \textquotedblleft Modified sphere decoding for
sparse code multiple access,\textquotedblright{} \textit{IEEE Commun.
Lett.}, vol. 22, no. 8, pp. 1544--1547, Aug. 2018.

\bibitem{key-7}\label{R.-Y.-Mesleh}R. Y. Mesleh, H. Haas, S. Sinanovic,
C. W. Ahn, and S. Yun, \textquotedblleft Spatial modulation,\textquotedblright{}
\textit{IEEE Trans. Veh. Technol.}, vol. 57, no. 4, pp. 2228--2241,
Jul. 2008.

\bibitem{key-9}\label{A.-Younis,-N.}A. Younis, N. Serafimovski,
R. Mesleh, and H. Haas, \textquotedblleft Generalised spatial modulation,\textquotedblright{}
in \textit{Proc. Forty Fourth Asilomar Conf. Signals, Syst., Comput.
(ASILOMAR)}, Nov. 2010, pp. 1498--1502.

\bibitem{key-5}\label{M.-Renzo,-H.}M. Renzo, H. Haas, A. Ghrayeb,
S. Sugiura, and L. Hanzo, \textquotedblleft Spatial modulation for
generalized MIMO: Challenges, opportunities and implementation,\textquotedblright{}
\textit{Proc. IEEE}, vol. 102, no. 1, pp. 56--103, Jan. 2014.

\bibitem{key-8-1}\label{E.-Basar,-=002018=002018Index}E. Basar,
\textquoteleft \textquoteleft Index modulation techniques for 5G wireless
networks,\textquoteright \textquoteright{} \textit{IEEE Commun. Mag.,}
vol. 54, no. 7, pp. 168--175, Jul. 2016.

\bibitem{key-1}\label{T.-Mao,-Q.}T. Mao, Q. Wang, Z. Wang, and S.
Chen, ``Novel index modulation techniques: A survey,'' \textit{IEEE
Commun. Surveys Tuts.}, vol. 21, no. 1, pp. 315-348, 1st quarter 2019.

\bibitem{key-2}\label{M.-Wen-et}M. Wen et al., ``A survey on spatial
modulation in emerging wireless systems: Research progresses and applications,''
\textit{IEEE J. Sel. Areas Commun.}, vol. 37, no. 9, pp. 1949-1972,
Sep. 2019.

\bibitem{key-6}\label{A.-Younis,-R. GLOBECOM}A. Younis, R. Mesleh,
H. Haas, and P. M. Grant, \textquotedblleft Reduced complexity sphere
decoder for spatial modulation detection receivers,\textquotedblright{}
in \textit{Proc. IEEE GLOBECOM}, 2010, pp. 1--5.

\bibitem{key-7}\label{A.-Younis,-M. ICC}A. Younis, M. Di Renzo,
R. Mesleh, and H. Haas, \textquotedblleft Sphere decoding for spatial
modulation,\textquotedblright{} in \textit{Proc. 2011 IEEE Int. Conf.
Commun.}, pp. 1--6.

\bibitem{key-8}\label{A.-Younis,-S. Trans}A. Younis, S. Sinanovic,
M. Di Renzo, R. Mesleh, and H. Haas, \textquotedblleft Generalised
sphere decoding for spatial modulation,\textquotedblright{} \textit{IEEE
Trans. Commun}., vol. 61, no. 7, pp. 2805--2815, July 2013.

\bibitem{key-13}\label{I.-Al-Nahhal,-O. QSM SD}I. Al-Nahhal, O.
A. Dobre, and S. Ikki, \textquotedblleft Quadrature spatial modulation
decoding complexity: Study and reduction,\textquotedblright{} \textit{IEEE
Wireless Commun. Lett.,} vol. 6, pp. 378-381, Jun. 2017.

\bibitem{key-14}\label{I.-Al-Nahhal,-O. VTC}I. Al-Nahhal, O. A.
Dobre, and S. Ikki, \textquotedblleft Low complexity decoders for
spatial and quadrature spatial modulations,\textquotedblright{} in
\textit{Proc. IEEE Veh. Technol. Conf. (VTC-Spring)}, 2018, pp. 1--5.

\bibitem{key-3}\label{I.-Al-Nahhal,-O.vtc20}I. Al-Nahhal, O. A.
Dobre, and S. Ikki, \textquotedblleft Reliable detection for spatial
modulation systems,\textquotedblright{} arXiv preprint arXiv:2006.05084.

\bibitem{key-15}\label{I.-Al-Nahhal,-E. JSAC}I. Al-Nahhal, E. Basar,
O. A. Dobre, and S. Ikki, \textquotedblleft Optimum low-complexity
decoder for spatial modulation,\textquotedblright{} \textit{IEEE J.
Sel. Areas Commun.}, vol. 37, no. 9, pp. 2001-2013, Jul. 2019.

\bibitem{key-10}\label{C.-Zhong,-X.}C. Zhong, X. Hu, X. Chen, D.
W. Ng, and Z. Zhang, ``Spatial modulation assisted multi-antenna
non-orthogonal multiple access,'' \textit{IEEE Wireless Commun}.\textit{
Lett}.\textit{,} vol. 25, no. 2, pp. 61-67, Apr. 2018.

\bibitem{key-11}\label{Y.-Liu,-L.}Y. Liu, L. L. Yang, and L. Hanzo,
\textquotedblleft Spatial modulation aided sparse code division multiple
access,\textquotedblright{} \textit{IEEE Trans. Wireless Commun.},
vol. 17, no. 3, pp. 1474--1487, Mar. 2018.

\bibitem{key-12}\label{Z.-Pan,-J.}Z. Pan, J. Luo, J. Lei, L. Wen,
and C. Tang, ``Uplink spatial modulation SCMA system,'' \textit{IEEE
Commun}.\textit{ Lett}.\textit{, }vol. 23, no. 1, pp. 184-187, Jan.
2019.

\bibitem{key-16}\label{I.-Al-Nahhal,-O. TVT}I. Al-Nahhal, O. A.
Dobre, E. Basar, and S. Ikki, \textquotedblleft Low-cost uplink sparse
code multiple access for spatial modulation,\textquotedblright{} \textit{IEEE
Trans. Veh. Technol.}, vol. 68, no. 9, pp. 9313-9317, Jul. 2019.

\bibitem{key-5}\label{L.-G.-Barbero}L. G. Barbero and J. S. Thompson,
\textquotedblleft Fixing the complexity of the sphere decoder for
MIMO detection,\textquotedblright{} \textit{IEEE Trans. Wireless Commun.,}
vol. 7, pp. 2131--2142, Jun. 2008.

\bibitem{key-1}\label{I.-Al-Nahhal,-A. comex}I. Al-Nahhal, A. Emran,
H. Kasem, A. B. Abd El-Rahman, O. Muta, and H. Furukawa, ``Flexible
fractional K-best sphere decoding for uncoded MIMO channels,'' \textit{IEICE
Communications Express}, vol., 4, pp. 20-25, Jan. 2015.

\bibitem{key-1}\label{I.-Al-Nahhal,-M.ModifiedZF}I. Al-Nahhal, M.
Alghoniemy, A. B. Abd El-Rahman, Z. Kawasaki, ``Modified zero forcing
decoder for ill-conditioned channels'' in \textit{Proc. }IFIP Wireless
Days (WD), Nov 2013, pp. 1--3.

\bibitem{key-1}\label{I.-Al-Nahhal,-M. Kbest ill condition}I. Al-Nahhal,
M. Alghoniemy, O. Muta, and A. B. A. El-Rahman, ``Reduced complexity
k-best sphere decoding algorithms for ill-conditioned MIMO channels,''
in \textit{Proc. IEEE Annu. Consum. Commun. Netw. Conf.}, pp. 183-187,
Jan. 2016.

\bibitem{key-1}\label{J.-Proakis,-Digital}J. Proakis, \textit{Digital
Communications}, 4th ed. New York, NY, USA: McGraw-Hill, 2000.
\end{thebibliography}
\end{document}